\documentclass[twocolumn]{jpsj2}
\def\i{{\rm i}}
\def\d{{\rm d}}
\def\e{{\rm e}}
\def\vector#1{{\mib{#1}}}
\def\va{{\vector a}}

\def\vk{{\vector k}}
\def\vK{{\vector K}}

\def\vq{{\vector q}}
\def\vQ{{\vector Q}}
\def\vr{{\vector r}}
\def\vR{{\vector R}}

\def\dps{\displaystyle}

\def\Tc{{T_{\rm c}}}

\def\hightc{{high-$T_{\rm c}$ }}

\def\TMTSFX{\mbox{${\rm (TMTSF)_2X}$}}
\def\TMTSFClO{\mbox{${\rm (TMTSF)_2ClO_4}$}}
\def\TMTSFPF{\mbox{${\rm (TMTSF)_2PF_6}$}}

\def\SrRuO{\mbox{${\rm Sr_2RuO_4}$}}

\def\hsp#1{\hspace{#1ex}}

\def\Tc{{T_{\rm c}}}

\def\lsim{\stackrel{{\textstyle<}}{\raisebox{-.75ex}{$\sim$}}}
\def\gsim{\stackrel{{\textstyle>}}{\raisebox{-.75ex}{$\sim$}}}

\def\dblesum#1#2{\sum_{\stackrel{\scriptstyle #1}{\scriptsize #2}}}

\def\kF{k_{{\rm F}}}

\def\omegaD{{\omega_{\rm D}}}

\def\eq.#1{eq.~(\ref{#1})}
\def\refeq#1{(\ref{#1})}
\def\pardif#1#2{\frac{\partial #1}{\partial #2}}

\newcommand\Equation[3]{
\begin{equation}\label{#1}\tag{#2}
#3
\end{equation}
}
\title{
Effects of Short-Range Correlations on the Coulomb Screening and \\
the Pairing Interactions in Electron-Phonon Systems \\ 
-- Triplet Pairing Mediated by Phonons -- 
}

\author{Hiroshi {\sc Shimahara}}
\def\runauthor{Hiroshi {\sc Shimahara}}

\inst{
Department of Quantum Matter Science, ADSM, Hiroshima University, 
Higashi-Hiroshima 739-8530, Japan
}

\recdate{March ~~, 2004}
\abst{
Effects of short-range correlations on the Coulomb screening, 
the phonons, and the pairing interactions 
are examined in electron-phonon systems. 
First, we derive a model Hamiltonian of Coulomb interactions 
which includes both the long-range part $v_{\vq}$ 
and the short-range part $U$. 
It is found from the expression of the dielectric function 
that the strong on-site correlations weaken the Coulomb screening. 
Secondly, we examine the screened phonons and 
the interaction mediated by phonons. 
In a consistent picture, 
we derive an expression of the effective interaction 
which includes 
(1)~the screened Coulomb interactions, 
(2)~the pairing interactions mediated by phonons, 
and 
(3)~the effective interactions mediated by spin and charge fluctuations. 
It is rewritten in a form of a summation of 
(a) the effective interactions of 
the pure Hubbard model without the long-range Coulomb interactions, 
and (b) the phonon-mediated interactions 
plus screened Coulomb interactions with corrections 
due to both $U$ and $v_{\vq}$. 
Thirdly, we derive an effective Hamiltonian analogous to 
the BCS Hamiltonian. 
Fourthly, for some typical values of parameters, 
we obtain the ground state phase diagrams. 
It is found that spin-triplet superconductivity mediated by phonons occurs 
when the short-range electron correlations are sufficiently strong, 
and the Coulomb screening is sufficiently weak. 
We estimate the orders of the transition temperatures 
when the triplet superconductivity occurs. 
The obtained values are realistic for 
existing candidates of the triplet superconductors 
as the order of the magnitudes. 
The possible relevance of the phonon-mediated interactions 
to the heavy fermion superconductor ${\rm UPt_3}$ 
and the layered superconductors 
such \TMTSFX {} and \SrRuO {} are briefly discussed. 
}

\kword{
phonon-mediated pairing interactions, 
Coulomb screening, 
electron-phonon system, 
overscreening, 
spin and charge fluctuations, 
exotic superconductors, 
spin-triplet pairing, 
heavy fermion superconductors, 
organic superconductors, 
ruthenate superconductors. 
}

\begin{document}
%% JPSJ %% 
\sloppy
%% JPSJ %% 
\maketitle

%% PR %% FOR TWO COLUMN ACTIVATE THE LINE BELOW
%% ]

%% PR %% \narrowtext

%% 1.
\section{Introduction}

It is widely known that overscreening of the repulsive Coulomb interactions 
between electrons by ions gives rise to attractive interactions 
for low frequencies $|\omega| \lesssim \omega_{\rm D}$~\cite{Sch83}. 
This dynamical effect is considered to be 
the origin of the pairing interactions mediated by phonons, 
which induce the superconductivity in metals. 
At the same time, the long-range Coulomb interactions are screened 
into short-range interactions, 
and consequently 
the energies of the optical phonons of short wavelengths 
are reduced into the accoustic type ${\omega}_{\vq} \sim |\vq|$.

On the other hand, the screening of electronic and ionic charges 
influences the momentum dependence 
of the attractive interactions through the dielectric function. 
If there were not the screening, 
the coupling constant between electrons and optical phonons 
diverges as $|\vq| \rightarrow 0$ 
due to the long-range nature of the Coulomb interactions. 
However, in actuality, 
such divergence is suppressed by the screening, 
but the coupling constant remains momentum dependent.

Owing to such momentum dependences of 
the electron-phonon coupling constant and the phonon energy, 
the resultant pairing interaction mediated by phonons 
depends on the momentum transfer. 
Hence, it includes the anisotropic components, 
when it is expanded in terms of appropriate basis functions. 
Usually, such anisotropic components are ignored, 
because they are considered to be much smaller 
than the isotropic component in most cases. 
In fact, the superconductivity in the ordinary metals, 
which are probably mediated by phonons, 
are considered to be of $s$-wave pairing.

However, as Foulkes and Gyorffy pointed out~\cite{Fou77}, 
when the short-range Coulomb interactions are strong enough 
to suppress isotropic pairing, 
anisotropic pairing can be induced by 
the subdominant anisotropic interaction. 
They have examined possibility of spin-triplet superconductivity 
by this mechanism in metals such as Rh, W, and Pd.

Abrikosov discussed \hightc cuprates, 
and proposed that such momentum dependence of 
the phonon-mediated pairing interaction could give rise to 
extended $s$-wave superconductivity~\cite{Abr94}, 
although it seems from some experiments 
that the cuprates are $d$-wave superconductors. 
Following his theory, 
Bouvier and Bok calculated the superconducting gap function 
and obtained anisotropic momentum dependence~\cite{Bou95}. 
Friedel and Kohmoto, and Chang, Friedel and Kohmoto 
examined the $d$-wave pairing interactions mediated by phonons 
in the cuprates~\cite{Fri00,Cha00}. 
The present author and Kohmoto have proposed~\cite{Shi02a} 
that in the presence of coexisting ferromagnetic long-range order, 
a subdominant triplet state occurs 
after the singlet states are suppressed by the strong exchange field. 
We have discussed a possibility of this mechanism 
in ${\rm UGe_2}$~\cite{Sax00,Kir01}.

It is easily verified 
that the momentum dependence of the interaction is stronger 
for weaker screening. 
For example, in the Thomas-Fermi approximation, 
the coupling constant has a peak with the width of 
the inverse screening length in the momentum space. 
Therefore, in the interactions, 
the ratios of the anisotropic components to the isotropic component 
increase~\cite{NoteWSCR} 
when the screening becomes weaker. 
There are some possible factors which weaken the screening effect, 
such as crystal structures, 
low density of charge carriers, 
and low density of states. 
The present author and Kohmoto have examined layer structures as 
the weakening mechanism~\cite{Shi02b}, 
because some candidates of the triplet superconductors 
have layer structures as we shall discuss below. 
It was found that when the layer interval increases, 
the screening becomes weaker, 
and subdominant triplet pairing interaction could more easily 
overcome the dominant $s$-wave pairing interaction 
with an asist of the on-site Coulomb repulsion.

It should be noted~\cite{Shi02a,Shi02b} that 
the on-site Coulomb interaction is specific 
to the electron systems in solid, 
and does not appear in the electron gas model in the continuum space. 
The magnitude of the on-site Coulomb energy $U$ depends on 
the profile of the Wannier functions. 
Since such an on-site interaction is constant 
in the crystal momentum space, 
it only suppresses isotropic pairing, 
and consequently it favors subdominant anisotropic pairing~\cite{Shi02b}.

With respect to application, 
we are motivated by recently discovered exotic superconductors, 
such as \SrRuO, organic superconductors, 
and heavy fermion superconductors. 
In the \SrRuO {} superconductor, 
spin-triplet pairing has been suggested 
by a Knight shift measurement~\cite{Ish98} 
and a $\mu SR$ experiment~\cite{Luk98}. 
Many theories have been proposed on this material, 
and there are some controversies 
especially on the momentum dependence of 
the superconducting gap function~\cite{Mac03}. 
Particularly, 
some nonphonon mechanisms have been considered~\cite{Mac03}, 
but there has not been any evidence at present. 
On the other hand, the \SrRuO {} superconductor exhibited 
the isotope effect coefficients 
with unusual dependence on 
the impurity (or oxygen defect) concentration 
and a reverse isotope effect 
in clean samples~\cite{Mao01}. 
The formar can be explained by assuming 
internal transition of superconductivity~\cite{Shi03a}, 
if the phonon-mediated pairing interaction exists. 
For the latter result, it has been shown~\cite{Shi03b} 
that within the weak coupling theory, 
the reverse isotope effect indicates that 
the Coulomb interaction between electrons, in total effect, 
works as repulsive interactions against the superconductivity, 
unless the system is in the very vicinity of any magnetic instability. 
At present, any pure nonphonon theory could not explain 
these experimental facts of the isotope effect.

Another candidate of the present mechanism of triplet superconductivity 
is the family of \TMTSFX {} compounds. 
From the experimental results at the early stage after the discovery 
of superconductivity in these compounds, 
including nuclear magnetic resonance (NMR)~\cite{Tak87,Has87} 
and the phase diagrams~\cite{Ish90}, 
$d$-wave superconductivity mediated by antiferromagnetic 
spin fluctuation has been studied 
in these compounds~\cite{Eme86,Shi89}. 
However, recent experimental studies suggest 
spin-triplet superconductivity in 
\TMTSFPF~\cite{Lee02} and \TMTSFClO~\cite{Oh04,Bel97}, 
although some of the experimental results have not been 
explained by triplet pairing~\cite{NoteORG}. 
We will examine this problem in a separate paper in details, 
applying the present mechanism of triplet pairing to 
these compounds~\cite{Sug04}. 
Only by the phonon mechanism, it might be difficult to reproduce 
the pressure dependence of the superconducting transition temperature 
$T_{\rm c}$ observed in these compounds~\cite{Ish90}. 
The pressure dependence may be explained by taking into account 
the contribution from the spin fluctuations to the pairing interactions. 
The pairing interactions mediated by the spin fluctuations include 
the attractive triplet components~\cite{Shi89,Shi00e}.

We also consider the heavy fermion superconductors, 
such as ${\rm UPt_3}$~\cite{Sig91,Tou96}, 
as candidates in which the pairing interactions mediated by phonons 
may contribute to spin-triplet pairing. 
Since the width of the effective band is extremely narrow, 
$s$-wave state could not avoid the on-site Coulomb repulsion 
by the retardation effect. 
If one use the Thomas-Fermi approximation in the estimation of 
the screening length, 
it becomes much smaller than the lattice constant. 
However, it is not justified for a phenomena of such a small length scale. 
In actuality, the dipole field due to the charge distributions 
in the unit cell could not decay so rapidly within 
the scale of the lattice constant. 
Hence, the electron-ion interaction is not necessarily limited 
on the each lattice site. 
Therefore, it may contribute to the anisotropic pairing interactions. 
Whatever the pairing mechanism is, 
since the heavy mass renormalization significantly reduces 
the effective coupling constant of the pairing interaction, 
the bare pairing interaction needs to be very large for the observed 
$T_{\rm c}$ to be reproduced.

In general, it is reasonable that 
the phonon-mediated pairing interaction 
could contribute to superconductivity 
to some extent in most cases, 
even when the resultant gap function is anisotropic. 
In fact, in almost all superconductors including exotic ones, 
non-vanishing isotope effects have been observed. 
It is reasonable to take them due to attractive contributions 
from phonons to the pairing interactions. 
It is likely that the momentum average like 
$\langle \gamma^{*}(\vk) V_{\rm ph}(\vk - \vk') \gamma(\vk') \rangle$ 
is negative, since the phonon-mediated pairing interaction 
$V_{\rm ph}(\vq)$ have a negative peak 
around $\vq = 0$ and $\left |\gamma(\vk) \right |^2 > 0$, 
where $\gamma(\vk)$ denotes the function that expresses 
the momentum dependence of the gap function 
$\Delta (\vk) \propto \gamma(\vk)$. 
Further, it is likely that the isotropic component is dominant, 
and the second and third-dominant ones are anisotropic 
and of odd and even parities in the momentum space, 
respectively.

In particular, in the spin-triplet superconductors, 
the phonon-mediated pairing interaction can be 
the main mechanism of the anisotropic superconductivity, 
although another mechanism might assist it at the same time~\cite{NoteCEX}. 
By an analogy to the superfluidity in liquid $^3$He,~\cite{Leg75} 
the paramagnon mechanism is sometimes considered to be responsible 
for the spin-triplet superconductivity. 
However, the situation is very different from the electron systems 
in the solids. 
The liquid $^3$He does not have lattice vibrations, 
and its transition temperature is of the order of 1mK~\cite{NoteHe3}. 
Therefore, we should be careful when we use the analogy.

In the anisotropic singlet superconductors, 
the third-dominant component of the phonon-mediated interaction 
may contribute to the superconductivity to some extent. 
However, we presume that it could not be the only origin 
of the pairing interaction. 
If the pairing interaction of nonphonon origin is negligible, 
the system should undergo a transition to 
triplet superconductivity at a higher temperature, 
since the third-dominant even-parity (spin-singlet) component 
is smaller than the second-dominant odd-parity (spin-triplet) component.

In this paper, we examine the effects of 
the short-range (on-site) correlations 
on the Coulomb screening, phonons, and the pairing interactions, 
taking into account the long-range Coulomb interactions 
at the same time. 
The main purpose of this paper is to devolop a general theory, 
although we are motivated by the superconductors discussed above. 
In particular, 
we derive a general effective Hamiltonian in a unified framework 
in which the short-range and long-range parts of 
the Coulomb interactions are treated consistently. 
The effective Hamiltonian includes 
(1)~the screened Coulomb interactions, 
(2)~the phonon-mediated interactions 
by the screened electron-phonon interactions, 
and (3)~the effective interactions mediated by 
spin and charge fluctuations. 
In appropriate limits, 
it is reduced to the effective Hamiltonians examined 
by many authors so far~\cite{Abr94,Bou95,Fri00,Cha00,Shi02a,Shi02b}. 
In particular, in our previous papers, 
the on-site Coulomb energy $U$ has been treated 
as a parameter independent of the phonon-mediated pairing interaction, 
but in practice it also modifies the pairing interactions. 
We will also clarify it in this paper.

In \S2, we examine the model of the electron-phonon system. 
Starting from a basic model of the electron-phonon system, 
we derive a Hamiltonian which includes 
both the short-range part $U$ and the long-range part $v_{\vq}$ 
of the Coulomb interactions. 
In \S3, we apply the random phase approximation (RPA) 
to the Hamiltonian obtained in \S2, 
and examine the screening effects 
on the Coulomb interactions and phonons. 
The approximation is reduced to that in the electron gas model 
if we put $U = 0$ and $v_{\vq} \ne 0$, 
while that in the pure Hubbard model 
if we put $U \ne 0$ and $v_{\vq} = 0$. 
In \S4, we examine the two-particle vertex part 
which contributes to superconductivity. 
We derive the general form which includes corrections 
due to $U$ as mentioned above. 
In \S5, we derive an effective interaction 
within the weak coupling theory from the results of \S4. 
We examine anisotropic superconductivity 
on the basis of the effective interaction. 
Expressions of the transition temperature and 
the isotope effect coefficient are derived. 
In \S 6, we apply the effective model to some typical cases 
to clarify essential aspects of the present mechanism. 
The phase diagrams and the transition temperatures are obtained. 
The last section \S 7 is devoted to summary and discussion.

%%% 2. 
\section{A Model of the Electron-Phonon System}

In this section, we derive a model that we will examine in this paper. 
We start with the general form of the coupled electron-phonon Hamiltonian 
\Equation{eq:electronphononHamiltonian}
{2.1}
{
     H = H_{0} + H_{\mbox{\tiny e-ph}} + H_{\rm C}
     }
with 
\Equation{eq:H0}
{2.2}
{
     H_{0} 
     = 
          \sum_{\vk \sigma} \epsilon_{\vk} 
                 c_{\vk \sigma}^{\dagger} c_{\vk \sigma}
        + \sum_{\vq \lambda} \Omega_{\vq \lambda} 
                 b_{\vq \lambda}^{\dagger} b_{\vq \lambda}
     }
\Equation{eq:Heph}
{2.3}
{
     H_{\mbox{\tiny e-ph}} 
     = 
          \sum_{\vk \vq \sigma \lambda} 
               M_{\vk \vq \lambda}
                 c_{\vk+\vq \sigma}^{\dagger} c_{\vk \sigma}
                 (    b_{\vq \lambda}
                    + b_{-\vq \lambda}^{\dagger} ) 
     }
\Equation{eq:HCoulomb}
{2.4}
{
     H_{\rm C}
     = 
          \frac{1}{2} 
          \dblesum{\vk_1 \cdots \vk_4}{\sigma \sigma'}
          V_{\vk_1 \vk_2 \vk_3 \vk_4}^{\sigma \sigma'} 
          c_{\vk_1 \sigma}^{\dagger} c_{\vk_2 \sigma}
          c_{\vk_3 \sigma'}^{\dagger} c_{\vk_4 \sigma'} . 
     }
Here, $c_{\vk \sigma}$ denotes the electron operator of the Bloch state 
with the crystal momentum $\vk$ and the spin $\sigma$ 
in the relevant electron band, 
while $b_{\vq \lambda}$ donotes the bare phonon operator of the phonon 
with the momentum $\vq$ and the phonon-mode $\lambda$. 
Since we consider the longitudinal optical phonons in this paper, 
we omit the phonon-mode suffix $\lambda$ and 
the summation over it for simplicity. 
Thus, we put $\Omega_{\vq \lambda} = \Omega_{\rm p}$, 
where $\Omega_{\rm p}$ denotes the ionic plasma frequency. 
We consider the positive background charge of the atoms within 
the jellium model~\cite{Sch83}. 
Therefore, we put the matrix element of the electron-phonon 
interaction $M_{\vk \vq \lambda} = M_{\vq}$ which satisfies 
\Equation{eq:ephcoupling}
{2.5}
{
     \frac{2 M_{\vq}^2}{\hbar \Omega_{\rm p}}
     = \frac{4 \pi e^2}{V_{\rm cell} \vq^2} , 
     }
where $V_{\rm cell}$ denotes the unit cell volume. 
The present model might not be very accurate, 
but since we need an order estimation at most for our purpose, 
we use this model in this paper for simplicity.

The coupling constant $V_{\vk_1 \vk_2 \vk_3 \vk_4}^{\sigma \sigma'}$ 
is expressed as 
\Equation{eq:Vinpsi}
{2.6}
{
     \begin{split}
     V_{\vk_1 \vk_2 \vk_3 \vk_4}^{\sigma \sigma'}
     &=   \int \d^3 \vr \, \d^3 \vr' \, 
          \psi_{\vk_1 \sigma}^{*}(\vr) \, \psi_{\vk_2 \sigma}(\vr) \\
     & \hsp{4}
          \times 
          \frac{e^2}{|\vr - \vr'|} 
          \psi_{\vk_3 \sigma'}^{*}(\vr') \, \psi_{\vk_4 \sigma'}(\vr') 
     \end{split}
     }
in terms of the Bloch wave functions 
\Equation{eq:psiBloch}
{2.7}
{
     \psi_{\vk}(\vr) = \e^{{\rm i} \vk \cdot \vr} u_{\vk}(\vr) , 
     }
where $u_{\vk}(\vr)$ satisfies the periodicity condition 
$u_{\vk}(\vr + \vR) = u_{\vk}(\vr)$ for any lattice vector $\vR$, 
according to the Bloch's theorem.

%%% 2.1 
\subsection{An expression in the tight-binding model} 

Now, we rewrite the Coulomb interaction $H_{\rm C}$ 
in the tight-binding model. 
When the tight-binding approximation is valid 
and the electron wave function has a large amplitude 
only near the atomic site, 
we may assume that 
\Equation{eq:psiassumed}
{2.8}
{
     \psi_{\vk} (\vr) 
     = \frac{1}{\sqrt{N}} 
          \sum_{\vR} \e^{ {\rm i} \vk \cdot \vR } \phi( \vr - \vR ) , 
     }
where $\phi(\vr - \vR)$ denotes a localized orbital wave function 
near the atomic site at $\vR$. 
This approximation is equivalent to putting 
\Equation{eq:ukrassumed}
{2.9}
{
     u_{\vk}(\vr) 
     = \frac{1}{\sqrt{N}} 
          \sum_{\vR} \e^{ - {\rm i} \vk \cdot (\vr - \vR) } 
                    \phi(\vr - \vR) . 
     }
Since we may put $\vr \approx \vR$ in \eq.{eq:ukrassumed}, 
it is equivalent also to 
$u_{\vk}(\vr) \approx N^{-1/2} \sum_{\vR} \phi(\vr - \vR)$. 
If we define the Wannier function 
\Equation{eq:Wannierdef}
{2.10}
{
     w(\vr - \vR) 
     = \frac{1}{\sqrt{N}} 
          \sum_{\vk} \e^{ - {\rm i} \vk \cdot \vR } 
            \psi_{\vk} (\vr) , 
     }
we obtain $w(\vr) = \phi(\vr)$ from \eq.{eq:psiassumed}.

Within this approximation, we obtain 
\Equation{eq:Vinphi}
{2.11}
{
     \begin{split}
     V_{\vk_1 \vk_2 \vk_3 \vk_4}^{\sigma \sigma'}
     &=   
          \frac{1}{N^{2}} 
          \sum_{\vR \vR'} 
          \int \d^3 \vr \, \d^3 \vr' \, 
          \left | \phi(\vr - \vR) \right |^2 
     \\ 
     & \hsp{-10}
          \times 
          \left | \phi(\vr' - \vR') \right |^2 
          \e^{ - {\rm i} (\vk_1 - \vk_2) \cdot \vR }
          \frac{e^2}{|\vr - \vr'|} 
          \e^{ - {\rm i} (\vk_3 - \vk_4) \cdot \vR' } , 
     \end{split}
     }
where we have omitted terms including the factor of the form 
$\phi(\vr - \vR_1) \phi(\vr - \vR_2)$ with $\vR_1 \ne \vR_2$, 
since it is small from the assumption mentioned above. 
In the integrand of \eq.{eq:Vinphi}, 
we could replace $\vr$ and $\vr'$ in $e^2/|\vr - \vr'|$ 
with $\vR$ and $\vR'$, respectively, 
due to the factors of $|\phi(\vr - \vR)|^2|\phi(\vr' - \vR')|^2$, 
except when $\vR = \vR'$. 
We could not make this replacement when $\vR = \vR'$, 
because $e^2/|\vr - \vr'|$ diverges when $\vr \rightarrow \vr'$. 
Therefore, we obtain 
\Equation{eq:VinUv}
{2.12}
{
     V_{\vk_1 \vk_2 \vk_3 \vk_4}^{\sigma \sigma'}
     = (U {\bar \delta}_{\sigma \sigma'} + v_{\vk_1 - \vk_2}) 
          \delta_{\vk_1 - \vk_2, \vk_3 - \vk_4} , 
     }
with ${\bar \delta}_{\sigma \sigma'} = 1 - \delta_{\sigma \sigma'}$, 
where we have defined 
\Equation{eq:Uexpression}
{2.13}
{
     U = 
          \int \d^3 \vr \d^3 \vr' \, \, 
          | \phi(\vr) |^2 \frac{e^2}{|\vr - \vr'|} | \phi(\vr') |^2 , 
     }
and 
\Equation{eq:vqexpression}
{2.14}
{
     v_{\vq} = 
          \sum_{\mib{\rho} \ne {\vector 0}} 
               \frac{e^2}{|\mib{\rho}|} 
                    \e^{- \i \vq \cdot \mib{\rho}} 
     }
for $\vq \ne 0$. 
It is obvious that the on-site Coulomb energy $U$ 
depends on the profile of the Wannier function, 
which expresses the charge distribution near the atomic site. 
Here, we put $v_{\vq = 0} = 0$, 
taking into account the charge neutrality in the present jellium model. 
We call the on-site $U$ and the inter-site $v_{\vq}$ 
the short-range and long-range parts of the Coulomb interaction, 
respectively. 
If we put $v_{\vq} = 0$, the present model is reduced to 
the pure Hubbard model on a lattice, 
while if we put $U = 0$, it is reduced to the electron gas model 
in the continuum space except that the momenta are replaced 
with the crystal momenta.

It is convenient to write \eq.{eq:HCoulomb} with matrices as 
\Equation{eq:HCmatrix}
{2.15}
{
     \begin{split}
     H_{\rm C}
     & = 
          \frac{1}{2N} \sum_{\vq \sigma \sigma'} 
          n_{\vq \sigma}
          \Bigl [ 
                 v_{\vq} {\hat \sigma}_0 
               + (v_{\vq} + U) {\hat \sigma}_1 
          \Bigr ]_{\sigma \sigma'} 
          n_{\vq \sigma'}
     \\
     & = 
          \frac{1}{2N} \sum_{\vq}
          \left ( \hsp{-0.5}
            \begin{array}{cc}
              n_{\vq \uparrow}  &  \hsp{-1.5}  n_{\vq \downarrow} 
            \end{array}
          \hsp{-0.5} \right ) 
          {\hat V}_{\vq} 
          \left ( \hsp{-0.5}
            \begin{array}{c}
              n_{\vq \uparrow}   \\
              n_{\vq \downarrow} 
            \end{array}
          \hsp{-0.5} \right ) , 
     \end{split}
     }
where we have introduced the Pauli matrices 
and the unit matrix, 
${\hat \sigma}_1$, ${\hat \sigma}_2$, ${\hat \sigma}_3$, 
and ${\hat \sigma}_0$, 
respectively. 
Here, we have defined 
$n_{\vq \sigma} = \sum_{\vk} c_{\vk\sigma}^{\dagger} c_{\vk + \vq \sigma}$ 
and 
\Equation{eq:Vmatrix}
{2.16}
{
     {\hat V}_{\vq} \equiv 
          \left ( \hsp{-0.5}
            \begin{array}{cc}
              v_{\vq}       & \hsp{-1.5} v_{\vq} + U  \\
              v_{\vq} + U   & \hsp{-1.5} v_{\vq} 
            \end{array}
          \hsp{-0.5} \right ) . 
}
We could also rewrite \eq.{eq:HCoulomb} as 
\Equation{eq:chargespin}
{2.17}
{
     H_{\rm C}
     = 
          \frac{1}{N} \sum_{\vq}
          \Bigl [ 
          \frac{1}{2} 
          ( v_{\vq} + \frac{U}{2} ) 
            n_{\vq} n_{-\vq} 
            - U S_{\vq}^{z} S_{-\vq}^{z} 
          \Bigr ]
     }
with $n_{\vq} = n_{\vq \uparrow} + n_{\vq \downarrow}$ 
and $S_{\vq}^{z} = (n_{\vq \uparrow} - n_{\vq \downarrow})/2$.

%%% 2.2 
\subsection{Approximation for the long-range part}

Now, we examine the long-range parts $v_{\vq}$ in detail. 
For small $\vq \ne {\vector 0}$, we could approximate $v_{\vq}$ by 
\Equation{eq:vqforsmallq}
{2.18}
{
     v_{\vq} \approx \int \frac{e^2}{|\mib{\rho}|} 
          \e^{ - \i \vq \cdot \mib{\rho} }
          \frac{\d^3 \mib{\rho}}{V_{\rm cell}}
     = \frac{4 \pi e^2}{V_{\rm cell} |\vq|^2} . 
     }
Here, we note that a periodicity relation $v_{\vq + \vK} = v_{\vq}$ 
holds in the original expression \eq.{eq:vqexpression}, 
where $\vK$ denotes any reciprocal lattice vector, 
whereas it does not hold in the approximate expression 
\eq.{eq:vqforsmallq}. 
Therefore, if one uses \eq.{eq:vqforsmallq} 
in calculating the gap function $\Delta(\vk)$, 
the reciprocal lattice periodicity 
$\Delta(\vk) = \Delta(\vk + \vK)$ of the gap function 
is broken.

Therefore, we note that the original expression of $v_{\vq}$, 
\eq.{eq:vqexpression}, is large near $\vq = \vK$, 
not only near $\vq = 0$, for the lattice periodicity. 
In the gap equation, the momentum $\vq$ corresponds 
to the momentum transfer $\vk - \vk'$, 
where $\vk$ and $\vk'$ denote the electron momenta 
near the Fermi surface. 
Even when $\vk$ and $\vk'$ are in the first Brillouin zone, 
$\vq - \vK = \vk - \vk' - \vK$ can be small, 
for example when the Fermi surface is open. 
For such $\vk$ and $\vk'$, 
we should use the approximate expression 
$v_{\vq} = 4 \pi e^2/V_{\rm cell}|\vq - \vK|^2$, 
which is very different from \eq.{eq:vqforsmallq} 
near $\vq = \vK$. 
Therefore, a more appropriate expression 
of \eq.{eq:vqexpression} is 
\Equation{eq:vqappoxcorrect}
{2.19}
{
     v_{\vq} \approx \max_{\vK} 
     \Bigl [
     \frac{4 \pi e^2}{V_{\rm cell}|\vq - \vK|^2} 
     \Bigr ] 
     }
rather than \eq.{eq:vqforsmallq}. 
This expression is appropriate not only near $\vq = 0$, 
but also near $\vq = \vK$. 
Near the Brillouin zone boundary of $\vq$, 
the derivative of the right-hand-side of 
\eq.{eq:vqappoxcorrect} jumps, 
but since $v_{\vq}$ is very small there, 
it does not cause any difficulty in practice.

\begin{figure}
%% JPSJ %% When you do not use epsf, activate the next line. 
%% \figureheight{7cm}
%% 
%% PR %% \begin{figure}[htb]
\begin{center}
%% FOR TWO COLUMN 
%% \leavevmode \epsfxsize=6cm  
%% 
%% FOR ONE COLUMN 
%% \leavevmode \epsfxsize=10cm  
%% JPSJ2 %% 
\includegraphics[width=7.0cm]{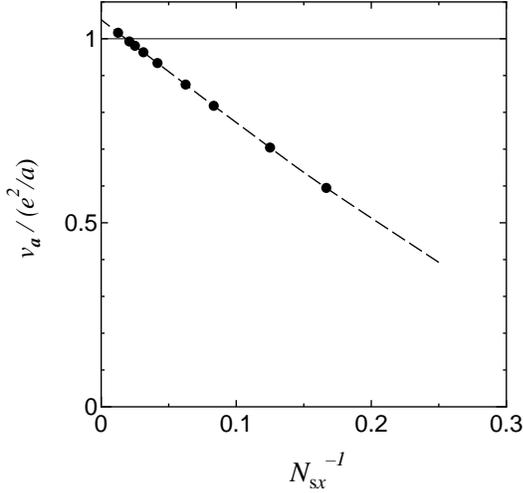}
%% PR %% \leavevmode \epsfxsize=7cm  
%% PR %% \epsfbox{fig01.eps}
\end{center}
\caption{
Estimation of the nearest neighbor Coulomb interaction $v_{\va}$ 
calculated with the approximate equation~\refeq{eq:vqappoxcorrect}
in the finite size system with the linear dimension $N_{sx}$. 
The broken line is the guide for eyes. 
The thin solid line shows the exact value $v_{\va} = 1$. 
} 
\label{fig:nnCoulomb}
\end{figure}

Let us examine the error of the above approximation. 
Since \eq.{eq:vqappoxcorrect} becomes exact relation 
for long wavelengths, 
it is expected that the error of \eq.{eq:vqappoxcorrect} 
is larger for shorter wavelengths. 
Hence, the nearest neighbor Coulomb interaction 
\Equation{eq:vaintegral}
{2.20}
{
     v_{\va} = N^{-1} \sum_{\vq} v_{\vq} \e^{{\rm i} q_x a} 
     }
would have the largest error, 
where $\va = (a,0,0)$ denotes one of the unit lattice vectors, 
since the on-site Coulomb interaction $U$ 
is not included in the definition of $v_{\vq}$, 
\eq.{eq:vqexpression}. 
Figure~\ref{fig:nnCoulomb} shows the result of \eq.{eq:vaintegral} 
calculated with the approximate equation~\refeq{eq:vqappoxcorrect} 
in the electron system on the cubic lattice 
with the number of the lattice sites $N = N_{{\rm s}x}^3$, 
where $N_{{\rm s}x}$ denotes the number of the lattice sites 
in one direction. 
It is found that as the system size increases, 
the value of $v_{\va}$ estimated by \eq.{eq:vqappoxcorrect} 
approach a value $\approx 1.05 \times e^2/a$, 
which is only $5\%$ larger than the exact value. 
Therefore, the present approximation does not cost major error 
even in the short wavelength behavior.

%%%%%%%%%%%%%%%%%%%%%%%%%%%%%%%%%%%%%%%%%%%%%%%%%%%%%%%%%%%%%%%%%%%%%%%
%%  Fig.1                                                            %%
%%%%%%%%%%%%%%%%%%%%%%%%%%%%%%%%%%%%%%%%%%%%%%%%%%%%%%%%%%%%%%%%%%%%%%%
%% JPSJ %% 
%%%%%%%%%%%%%%%%%%%%%%%%%%%%%%%%%%%%%%%%%%%%%%%%%%%%%%%%%%%%%%%%%%%%%%%

%%% 3. 
\section{Coulomb Screening}

In this section, 
we examine the screening effects on 
the Coulomb interactions between two electrons, 
the electron-phonon interactions, and 
the phonon propagators. 
For convenience, we define appropriately scaled functions 
of the charge and spin fluctuations in the pure Hubbard model 
{\it i.e.}, the model with $U \ne 0$ and $v_{\vq} = 0$, 
in the RPA by 
\Equation{eq:susceptibilityHubbard}
{3.1}
{
     \begin{split}
     \chi_{\rm c}^{(0)}(q) 
     & = \frac{2 \chi_0(q)}{1 + U \chi_0(q)} 
     \\
     \chi_{\rm s}^{(0)}(q) 
     & = \frac{1}{2} \frac{\chi_0(q)}{1 - U \chi_0(q)} , 
     \end{split}
     }
where we have defined 
\Equation{eq:chi0}
{3.2}
{
     \begin{split}
     \chi_0(q) 
     & = 
     - T \sum_{n} N^{-1} \sum_{\vk} 
          G_{\sigma}^{(0)}(k + q) \, G_{\sigma}^{(0)}(k) \\
     & = 
     N^{-1} \sum_{\vk} 
     \frac{f(\xi_{\vk}) - f(\xi_{\vk + \vq})}
          {{\rm i} \nu_m + \xi_{\vk + \vq} - \xi_{\vk} } 
     \end{split}
     } 
with the unperturbed Green's function $G_{\sigma}^{(0)}$. 
Here, we have introduced the four-momentum notation 
such as $q = (\vq,{\rm i} \nu_m)$ 
and $k = (\vk, {\rm i} \omega_n)$.

%%% 3.1 
\subsection{Fluctuations and Screened Coulomb Interactions}

In this subsection, we derive an expression of 
the screened Coulomb interactions in the presence of 
both the short and long-range parts of the Coulomb interactions, 
$U$ and $v_{\vq}$.

\begin{figure}
%% JPSJ %% When you do not use epsf, activate the next line. 
%% \figureheight{7cm}
%% 
%% PR %% \begin{figure}[htb]
\begin{center}
%% FOR TWO COLUMN 
%% \leavevmode \epsfxsize=6cm  
%% 
%% FOR ONE COLUMN 
%% \leavevmode \epsfxsize=10cm  
%% JPSJ2 %% 
\vspace{2\baselineskip}
\includegraphics[width=4.0cm]{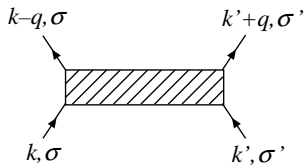}
\vspace{\baselineskip}
%% PR %% \leavevmode \epsfxsize=7cm  
%% PR %% \epsfbox{fig02.eps}
\end{center}
\caption{Assignment of the momenta and spins 
in the two body vertex part $\Gamma(k,k',q)$. 
} 
\label{fig:Gammamomass}
\end{figure}

\begin{figure}
%% JPSJ %% When you do not use epsf, activate the next line. 
%% \figureheight{7cm}
%% 
%% PR %% \begin{figure}[htb]
\begin{center}
%% FOR TWO COLUMN 
%% \leavevmode \epsfxsize=6cm  
%% 
%% FOR ONE COLUMN 
%% \leavevmode \epsfxsize=10cm  
%% JPSJ2 %% 
\vspace{2\baselineskip}
\includegraphics[width=5.0cm]{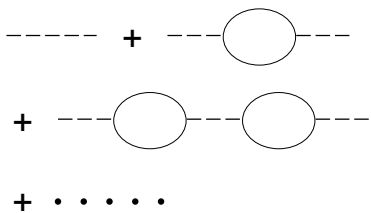}
\vspace{\baselineskip}
%% PR %% \leavevmode \epsfxsize=7cm  
%% PR %% \epsfbox{fig03.eps}
\end{center}
\caption{
Summation of the ring diagrams. 
The solid and broken lines denote the electron Green's function 
and the Coulomb interaction \eq.{eq:VinUv}. 
} 
\label{fig:RPA_ring}
\end{figure}

First, we consider the ring diagrams by the RPA 
as depicted in Fig.~\ref{fig:RPA_ring}, 
where the momenta and spins are assigned as shown 
in Fig.~\ref{fig:Gammamomass}. 
Since the dashed lines of the Coulomb interactions 
have the same momentum $\vq$ in those diagrams, 
they become divergently large at the same time 
for small $|\vq|$. 
Therefore, this series gives a major contribution 
for long wavelengths.

The vertex due to the ring diagrams is obtained as 
\Equation{eq:Vring}
{3.3}
{
     {\hat \Gamma}^{\rm ring}(k,k',q) 
     = 
          {\hat V}_{\vq} 
          \Bigl [ 
               {\hat \sigma}_0 + \chi_0(q) \, {\hat V}_{\vq} 
          \Bigr ]^{-1} 
     }
in the matrix form in the spin space. 
Equation \refeq{eq:Vring} is rewritten as 
\Equation{eq:Vringrewritten}
{3.4}
{
     \begin{split}
     {\hat \Gamma}^{\rm ring}(k,k',q) 
     & = 
     \frac{1}{2} 
     \frac{2 v_{\vq} + U } 
          { 1 + (2 v_{\vq} + U) \chi_0(q) } 
          \left ( \hsp{-0.5}
            \begin{array}{cc}
              1 & \hsp{-1.5} 1 \\
              1 & \hsp{-1.5} 1 
            \end{array}
          \hsp{-0.5} \right ) 
     \\ 
     & \hsp{3} 
     + 
     \frac{1}{2} 
     \frac{- U } 
          { 1 - U \chi_0(q) } 
          \left ( \hsp{-1}
            \begin{array}{cc}
              ~ 1 & \hsp{-1.5} - 1 \\
              - 1 & \hsp{-1.5} ~ 1 
            \end{array}
          \hsp{-1} \right ) , 
     \end{split}
     }
and we could put the elements in the form 
\Equation{eq:Vringelements}
{3.5}
{
     \Gamma^{\rm ring}_{\sigma \sigma'}(k,k',q) 
     = \Gamma_{\rm c}(q) + \Gamma_{\rm s}^{zz}(q) \sigma \sigma' . 
     }
In the static approximation ${\rm i} \nu_m = 0$, 
we have the effective interaction by the ring diagrams 
\Equation{eq:HCring}
{3.6}
{
     \begin{split}
     H_{\rm C}^{\rm ring}
     & = 
     \frac{1}{N} 
     \sum_{\vq}
     \Bigl [
     \frac{1}{2} 
     \frac{v_{\vq} + U/2 }{ 1 + 2 (v_{\vq} + U/2) \chi_0 (\vq,0) } 
     n_{\vq} n_{-\vq} 
     \\ 
     & \hsp{12} 
       - \frac{ U }{ 1 - U \chi_0(\vq,0) } 
           S_{\vq}^{z} S_{-\vq}^{z} 
     \Bigr ] . 
     \end{split}
     }
Here, we note that the rotational symmetry in the spin space 
is not retained if we take only the ring diagrams~\cite{NoteRTI}.

\begin{figure}
%% JPSJ %% When you do not use epsf, activate the next line. 
%% \figureheight{7cm}
%% 
%% PR %% \begin{figure}[htb]
\begin{center}
%% FOR TWO COLUMN 
%% \leavevmode \epsfxsize=6cm  
%% 
%% FOR ONE COLUMN 
%% \leavevmode \epsfxsize=10cm  
%% JPSJ2 %% 
\vspace{2\baselineskip}
\includegraphics[width=4.5cm]{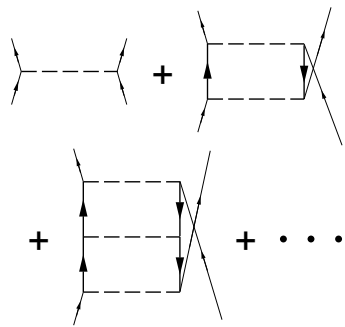}
%% \vspace{\baselineskip}
%% PR %% \leavevmode \epsfxsize=7cm  
%% PR %% \epsfbox{fig04.eps}
\end{center}
\caption{
Summation of the particle-hole ladder diagrams. 
The definitions of the lines are the same as 
those in Fig.~\ref{fig:RPA_ring}. 
The incoming and outgoing thin solid lines are to show how to apply 
the diagrams to the two-particle vertex 
in the self-energy and the superconductive gap equation. 
} 
\label{fig:RPA_ladder}
\end{figure}

In order to recover the rotational invariance, 
we need to take the particle-hole ladder diagrams 
depicted in Fig.~\ref{fig:RPA_ladder}, 
with respect to $U$. 
The second term in the braket of 
\eq.{eq:HCring} becomes divergently large 
for the momentum $\vq$ near the nesting vector $\vQ$, 
when the Fermi surface has a good nesting condition, 
as known in the Hubbard model~\cite{Sca86,Shi88}. 
It is also known that the contribution from the ladder diagrams 
in Fig.~\ref{fig:RPA_ladder} is of the same order 
as that from the ring diagrams.

In contrast, in the RPA in the pure electron gas model with $U = 0$, 
the ladder diagrams are ignored. 
Since the momentum of the interaction lines do not always 
coincide as those in the ring diagrams, 
the contribution from the ladder diagrams is smaller than 
that from the ring diagrams. 
Hence, we omit $v_{\vq}$ in the ladder diagrams also 
in the present model.

Thus, the contribution from the ladder diagrams 
is obtained as 
\Equation{eq:Vladder}
{3.7}
{
     \Gamma^{\rm lad}_{\sigma \sigma'}(k,k',q) 
     = 
          \frac{ - U }{ 1 - U \chi_0(k-k'-q) } 
          (1 - \delta_{\sigma \sigma'} ) . 
     }
In the static approximation, we have 
\Equation{eq:HClad}
{3.8}
{
     \begin{split}
     H_{\rm C}^{\rm lad}
     & = 
     - \frac{1}{2N} 
     \sum_{\vq}
       \frac{ U }{ 1 - U \chi_0 (\vq,0) } 
           (   S_{\vq}^{+} S_{-\vq}^{-} 
             + S_{\vq}^{-} S_{-\vq}^{+} ) , 
     \end{split}
     }
where we have defined 
$$
     S_{\vq}^{+} 
     = \sum_{\vk} c_{\vk        \uparrow}^{\dagger} 
                  c_{\vk + \vq  \downarrow}
     $$
and $S_{\vq}^{-} = (S_{\vq}^{+})^\dagger$. 
Adding eqs.~\refeq{eq:HCring} and~\refeq{eq:HClad}, 
we obtain a rotationally invariant effective interaction 
\Equation{eq:HCeff}
{3.9}
{
     \begin{split}
     {\tilde H}_{\rm C}
     & = 
     \frac{1}{N} 
     \sum_{\vq}
     \Bigl [
     \frac{1}{2}
     \frac{v_{\vq} + U/2}{ 1 + 2 (v_{\vq} + U/2) \chi_0 (\vq,0) } 
     n_{\vq} n_{-\vq} 
     \\ 
     & \hsp{5} 
       - \frac{ U }{ 1 - U \chi_0 (\vq,0) } 
           {\vector S}_{\vq} \cdot {\vector S}_{-\vq} 
     - U n_{\vq \uparrow} n_{-\vq \downarrow} 
     \Bigr ] , 
     \end{split}
     }
where the last term subtracts 
the double counted contribution 
of the first order diagrams 
in Figs.~\ref{fig:RPA_ring} and~\ref{fig:RPA_ladder}. 
From \eq.{eq:HCeff}, one could reproduce the vertex parts by the RPA 
both in the electron gas model and in the Hubbard model 
by taking appropriate limits, 
$U \rightarrow 0$ and $v_{\vq} \rightarrow 0$, respectively. 
When we use the effective Hamiltonian of \eq.{eq:HCeff} 
in the electron self-energy, we should subtract another 
double counted contribution from the second terms 
in Figs.~\ref{fig:RPA_ring} and~\ref{fig:RPA_ladder}~\cite{Shi88}.

Within the same approximation, 
the propagators for the charge and spin fluctuations 
are obtained as 
\Equation{eq:susceptibility}
{3.10}
{
     \begin{split}
     \chi_{\rm c}(q) 
     & = \frac{2 \chi_0(q)}{1 + ( 2 v_{\vq} + U) \chi_0(q)} 
     \\
     \chi_{\rm s}(q) 
     & = \frac{1}{2} \frac{\chi_0(q)}{1 - U \chi_0(q)} 
       = \chi_{\rm s}^{(0)}(q) . 
     \end{split}
     }
Hence, we can rewrite \eq.{eq:HCeff} as 
\Equation{eq:HCeffinsusceptibility}
{3.11}
{
     {\tilde H}_{\rm C}
     = 
     H_{\rm C} + H_{\rm cf} + H_{\rm sf} , 
     }
where 
\Equation{eq:HcfHsf}
{3.12}
{
     \begin{split}
     H_{\rm cf}
     & = 
     - \frac{1}{2 N} \sum_{\vq \ne 0}
           (v_{\vq} + \frac{U}{2})^2 \chi_{\rm c}(\vq,0) \, 
           n_{\vq} n_{-\vq} 
     \\ 
     H_{\rm sf}
     & = 
     - \frac{1}{2 N} \sum_{\vq \ne 0}
           4 U^2 \chi_{\rm s}(\vq,0) \, 
           {\vector S}_{\vq} \cdot {\vector S}_{-\vq} . 
     \end{split}
     }
The terms $H_{\rm cf}$ and $H_{\rm sf}$ 
can be regarded as the interactions mediated by 
the charge and spin fluctuations, respectively.

%%% 3.2 
\subsection{Dielectric function and the screening length}

The first terms of eqs.~\refeq{eq:Vringrewritten} 
and~\refeq{eq:HCeff} 
are the interactions between 
the charge degrees of freedom of electrons. 
Hence, we obtain an expression of the dielectric function 
\Equation{eq:diele}
{3.13}
{
     \kappa (q) = 1 + 2 (v_{\vq} + \frac{U}{2}) \chi_0(q) . 
     }
In the long wavelength limit $\vq \rightarrow 0$, 
since $v_{\vq} \gg U$, 
we obtain Thomas-Fermi dielectric constant 
\Equation{eq:dieleTF}
{3.14}
{
     \kappa (q) 
       = 1 + \frac{q_{\rm TF}^2}{\vq^2} \equiv \kappa_{\rm TF} (q) 
     }
with 
${
     {q_{\rm TF}}^2 = 8 \pi e^2 N(0) / V_{\rm cell} 
     }$, 
where $\lambda_{\rm TF} = 1/q_{\rm TF}$ and $N(0)$ denote 
the Thomas-Fermi screening length and the electron density of states 
at the Fermi energy per a spin and a site. 
For the short wavelength $|\vq| \gg q_{\rm s}$, 
we can omit $v_{\vq}$ compared to $U$, 
\Equation{eq:dieleshortwavelength}
{3.15}
{
     \kappa (q) = 1 + U \chi_0(q) \equiv \kappa_U(q) , 
     }
which is the result in the Hubbard model.

Equation \refeq{eq:diele} is written in the form, 
\Equation{eq:diele2}
{3.16}
{
     \kappa (q) 
       = \kappa_U(q) {\tilde \kappa}_{v}(q) , 
     }
where 
\Equation{eq:kappavtilde}
{3.17}
{
     \begin{split}
     {\tilde \kappa}_{v}(q) 
     & = 1 + v_{\vq} \chi_{\rm c}^{(0)}(q) 
       = 1 + \frac{{\tilde q}_s^2}{\vq^2} 
     \\
     {{\tilde q}_{\rm s}}^2 
     & = q_{\rm TF}^2   \frac{\chi_{\rm c}^{(0)}(q)}{2N(0)} 
       = \frac{q_{\rm s}^2}{\kappa_U(q)} 
     \\
     q_{\rm s}^2 & = q_{\rm TF}^2 \frac{\chi_0(q)}{N(0)}
     \end{split}
     }
with $\chi_{\rm c}^{(0)}$ defined in \eq.{eq:susceptibilityHubbard}. 
In the absence of $U$, 
in the system with the spherical Fermi surface, 
\eq.{eq:diele2} is reduced to the Lindhard's result of the RPA 
in the electron gas model.

For long wavelengths and in the static approximation, 
we have $\chi_0(q) \approx N(0)$ 
and $\kappa_U(q) \approx 1 + U N(0)$. 
Hence, we obtain the screening length 
\Equation{eq:screeninglength}
{3.18}
{
     {\tilde \lambda}_{\rm scr} 
          = 
          {\lambda}_{\rm TF} 
          \sqrt{\kappa_U(0)} 
          = 
          {\lambda}_{\rm TF} 
          \sqrt{1 + \mu_{\rm C}} 
     }
from \eq.{eq:kappavtilde}, 
where we have put 
\Equation{eq:muC}
{3.19}
{
     \mu_{\rm C} = U N(0) . 
     }
Therefore, it is found that the screening length is lengthened 
by a correlation effect due to the on-site Coulomb repulsion 
in comparison to the Thomas-Fermi screening length $\lambda_{\rm TF}$. 
It is plausible that the local electron density 
less easily deviates from the uniform value 
in the presence of strong correlations. 
In other words, the charge fluctuations are reduced 
($[\chi_{\rm c}^{(0)}]_{U \ne 0} < [\chi_{\rm c}^{(0)}]_{U = 0}$) 
and $\kappa_U > 1$, when $U > 0$. 
It is interesting that 
the short-range part of the Coulomb interactions affects 
the long-range behavior through a many body effect.

%%%%%%%%%%%%%%%%%%%%%%%%%%%%%%%%%%%%%%%%%%%%%%%%%%%%%%%%%%%%%%%%%%%%%%%
%%  Fig.2                                                            %%
%%%%%%%%%%%%%%%%%%%%%%%%%%%%%%%%%%%%%%%%%%%%%%%%%%%%%%%%%%%%%%%%%%%%%%%
%% JPSJ %% 
%%%%%%%%%%%%%%%%%%%%%%%%%%%%%%%%%%%%%%%%%%%%%%%%%%%%%%%%%%%%%%%%%%%%%%%

%%%%%%%%%%%%%%%%%%%%%%%%%%%%%%%%%%%%%%%%%%%%%%%%%%%%%%%%%%%%%%%%%%%%%%%
%%  Fig.3                                                            %%
%%%%%%%%%%%%%%%%%%%%%%%%%%%%%%%%%%%%%%%%%%%%%%%%%%%%%%%%%%%%%%%%%%%%%%%
%% JPSJ %% 
%%%%%%%%%%%%%%%%%%%%%%%%%%%%%%%%%%%%%%%%%%%%%%%%%%%%%%%%%%%%%%%%%%%%%%%

%%%%%%%%%%%%%%%%%%%%%%%%%%%%%%%%%%%%%%%%%%%%%%%%%%%%%%%%%%%%%%%%%%%%%%%
%%  Fig.4                                                            %%
%%%%%%%%%%%%%%%%%%%%%%%%%%%%%%%%%%%%%%%%%%%%%%%%%%%%%%%%%%%%%%%%%%%%%%%
%% JPSJ %% 
%%%%%%%%%%%%%%%%%%%%%%%%%%%%%%%%%%%%%%%%%%%%%%%%%%%%%%%%%%%%%%%%%%%%%%%

%%% 3.3 
\subsection{Corrections to the electron-phonon interactions}

\begin{figure}
%% JPSJ %% When you do not use epsf, activate the next line. 
%% \figureheight{7cm}
%% 
%% PR %% \begin{figure}[htb]
\begin{center}
%% FOR TWO COLUMN 
%% \leavevmode \epsfxsize=6cm  
%% 
%% FOR ONE COLUMN 
%% \leavevmode \epsfxsize=10cm  
%% JPSJ2 %% 
\vspace{2\baselineskip}
\includegraphics[width=6.5cm]{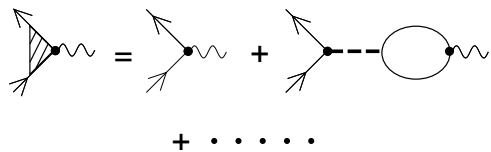}
\vspace{\baselineskip}
%% PR %% \leavevmode \epsfxsize=7cm  
%% PR %% \epsfbox{fig05.eps}
\end{center}
\caption{
Renormalization of the electron-phonon interaction. 
The solid lines denote the electron Green's functions, 
and the closed circle denotes the bare electron-phonon interaction. 
The thin solid and wavy lines are to show how to apply the diagrams 
to the electron and phonon Green's functions, respectively. 
The thick dashed line denotes the screened Coulomb interaction. 
} 
\label{fig:e-ph_renormalization}
\end{figure}

Now, let us examine renormalization of the electron-phonon 
interaction due to the Coulomb screening effect as depicted in 
Fig.~\ref{fig:e-ph_renormalization}. 
As the screened Coulomb interaction, we take the series of 
the ring diagrams shown in Fig.~\ref{fig:RPA_ring}. 
However, if we apply the ladder diagrams in thick dashed line 
in Fig.~\ref{fig:e-ph_renormalization}, 
the ring part consists of many electron lines. 
In such a ring, the integrand could be large only in a very small part 
of the phase space of the integral variables. 
Hence, we omit the ladder diagrams for simplicity. 
This simplification is also justified from the rotational invariance 
in the spin space. 
Since we could take the spin summation for the ring part 
in Fig.~\ref{fig:e-ph_renormalization}, 
the spin interaction in \eq.{eq:Vringrewritten} vanishes. 
Thus, the electron-phonon vertex part is not rotationally invariant 
if we retain the ladder diagrams. 
Therefore, we ignore them, 
and obtain the electron-phonon vertex part 
\Equation{eq:Mrenormalized}
{3.20}
{
     {\tilde M}(q) 
%%      = M_{\vq} \, 
%%           \left ( 
%%           1 - 
%%           \frac{2 v_{\vq} + U}{ 1 + ( 2 v_{\vq} + U ) \chi_0(q) } 
%%           \chi_0(q) 
%%           \right ) 
     = \frac{M_{\vq}}{\kappa(q)} . 
     }

%%%%%%%%%%%%%%%%%%%%%%%%%%%%%%%%%%%%%%%%%%%%%%%%%%%%%%%%%%%%%%%%%%%%%%%
%%  Fig.5                                                            %%
%%%%%%%%%%%%%%%%%%%%%%%%%%%%%%%%%%%%%%%%%%%%%%%%%%%%%%%%%%%%%%%%%%%%%%%
%% JPSJ %% 
%%%%%%%%%%%%%%%%%%%%%%%%%%%%%%%%%%%%%%%%%%%%%%%%%%%%%%%%%%%%%%%%%%%%%%%

%%% 3.4 
\subsection{Corrections to the phonons}

\begin{figure}
%% JPSJ %% When you do not use epsf, activate the next line. 
%% \figureheight{7cm}
%% 
%% PR %% \begin{figure}[htb]
\begin{center}
%% FOR TWO COLUMN 
%% \leavevmode \epsfxsize=6cm  
%% 
%% FOR ONE COLUMN 
%% \leavevmode \epsfxsize=10cm  
%% JPSJ2 %% 
\vspace{2\baselineskip}
\includegraphics[width=6.5cm]{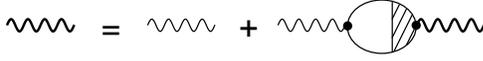}
\vspace{\baselineskip}
%% PR %% \leavevmode \epsfxsize=7cm  
%% PR %% \epsfbox{fig06.eps}
\end{center}
\caption{
The diagram equation for the dressed phonon Green's function. 
The thick and thin wavy lines denote the dressed and bare phonon 
Green's functions, respectively. 
The renormalized electron-phonon interaction is shown 
in Fig.~\ref{fig:e-ph_renormalization}. 
} 
\label{fig:phonon}
\end{figure}

\begin{figure}
%% JPSJ %% When you do not use epsf, activate the next line. 
%% \figureheight{7cm}
%% 
%% PR %% \begin{figure}[htb]
\begin{center}
%% FOR TWO COLUMN 
%% \leavevmode \epsfxsize=6cm  
%% 
%% FOR ONE COLUMN 
%% \leavevmode \epsfxsize=10cm  
%% JPSJ2 %% 
\vspace{2\baselineskip}
\includegraphics[width=6.5cm]{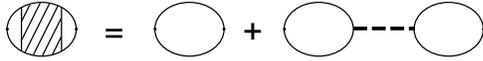}
\vspace{\baselineskip}
%% PR %% \leavevmode \epsfxsize=7cm  
%% PR %% \epsfbox{fig07.eps}
\end{center}
\caption{
The diagram equation for the polarization propagator. 
The solid line and thick dashed line denote the bare electron 
Green's function and the screened Coulomb interaction. 
} 
\label{fig:polarization}
\end{figure}

In this subsection, we examine the phonon Green's function. 
We consider the renormalization of the phonons 
due to the Coulomb screening 
through the electron-phonon interaction shown 
in Fig.~\ref{fig:e-ph_renormalization}. 
Then, we obtain the diagram equation shown in Fig.~\ref{fig:phonon}. 
This approximation is equivalent to that for the polarization 
propagator as shown in Fig.~\ref{fig:polarization}.

%%%%%%%%%%%%%%%%%%%%%%%%%%%%%%%%%%%%%%%%%%%%%%%%%%%%%%%%%%%%%%%%%%%%%%%
%%  Fig.6                                                            %%
%%%%%%%%%%%%%%%%%%%%%%%%%%%%%%%%%%%%%%%%%%%%%%%%%%%%%%%%%%%%%%%%%%%%%%%
%% JPSJ %% 
%%%%%%%%%%%%%%%%%%%%%%%%%%%%%%%%%%%%%%%%%%%%%%%%%%%%%%%%%%%%%%%%%%%%%%%

%%%%%%%%%%%%%%%%%%%%%%%%%%%%%%%%%%%%%%%%%%%%%%%%%%%%%%%%%%%%%%%%%%%%%%%
%%  Fig.7                                                            %%
%%%%%%%%%%%%%%%%%%%%%%%%%%%%%%%%%%%%%%%%%%%%%%%%%%%%%%%%%%%%%%%%%%%%%%%
%% JPSJ %% 
%%%%%%%%%%%%%%%%%%%%%%%%%%%%%%%%%%%%%%%%%%%%%%%%%%%%%%%%%%%%%%%%%%%%%%%

The phonon Green's function is defined by 
\Equation{eq:phononGf}
{3.21}
{
     \mathcal{D}(\vq,\tau)
     = 
     - \langle {\rm T}_{\tau} \left [ 
          \varphi_{\vq}(\tau) \varphi_{\vq}^{\dagger} \right ] 
          \rangle 
     }
with $\varphi_{\vq} = b_{\vq} + b_{-\vq}^{\dagger}$, 
where we have defined $A(\tau) = \e^{\tau H}A\e^{ - \tau H}$. 
The bare phonon Green's function $\mathcal{D}_0(\vq,\tau)$ 
is defined in a similar manner, 
and the Fourier transform is expressed as 
\Equation{eq:D0}
{3.22}
{
     \mathcal{D}_0 (\vq, {\rm i} \nu_m ) 
          = \frac{- 2 \Omega_{\rm p}}
                 {\nu_m^2  +  \Omega_{\rm p}^2} . 
     }

The polarization propagator of Fig.~\ref{fig:polarization} 
summed over spins is expressed as 
\Equation{eq:P}
{3.23}
{
     \begin{split}
     P(q) 
     & = 2 \chi_0(q) 
          - [\chi_0(q)]^2 \sum_{\sigma \sigma'} 
                (  \Gamma_{\rm c}(q) 
                 + \Gamma_{\rm s}^{zz} (q) \sigma \sigma' ) 
     \\
     & = 2 \chi_0(q) / \kappa(q) 
       \equiv \chi_{\rm c}(q) , 
     \end{split}
     }
where $\chi_{\rm c}(q)$ denotes the charge fluctuations. 
From Fig.~\ref{fig:phonon} 
with Fig.~\ref{fig:e-ph_renormalization} and \eq.{eq:Mrenormalized} 
or with Fig.~\ref{fig:polarization} and \eq.{eq:P}, 
we obtain 
\Equation{eq:Deq}
{3.24}
{
     \mathcal{D}(q) 
     = \mathcal{D}_0(q) 
       - \mathcal{D}_0(q) 
         \frac{2 M_{\vq}^2 \chi_0(q)}{\kappa(q)} 
         \mathcal{D}(q) , 
     }
and thus 
\Equation{eq:DresultRPAgeneral}
{3.25}
{
     \mathcal{D}^{-1} 
       = \mathcal{D}_0^{-1} + M_{\vq}^2 \chi_{\rm c}(q) . 
     }
Using eqs.~\refeq{eq:D0} and \refeq{eq:ephcoupling}, 
we obtain 
\Equation{eq:Dresult}
{3.26}
{
     \mathcal{D}^{-1} 
       = - \frac{1}{2 \Omega_{\rm p}} 
           \left [ 
              \nu_m^2 + [{\tilde \omega}(q)]^2 
           \right ] 
     }
with 
\Equation{eq:omegatilde}
{3.27}
{
     {\tilde \omega}(q) 
          = \Omega_{\rm p} 
               \sqrt{ \frac{\kappa_{U}(q)}{\kappa (q)} } 
          = \frac{\Omega_{\rm p}}{{\tilde \kappa}_{v}(q)} . 
     }
The renormalized phonon dispersion energy is obtained by 
solving 
${\tilde \omega}_{\vq} = {\tilde \omega}(\vq,{\tilde \omega}_{\vq})$, 
where the function ${\tilde \omega}(\vq,\omega)$ is 
obtained by the analitic continuation 
${\rm i} \nu_m \rightarrow \omega + {\rm i} \delta$ 
from ${\tilde \omega}(\vq,{\rm i}\nu_m)$.

In the long wavelength limit $\vq \rightarrow 0$, 
we obtain 
\Equation{eq:accoustic}
{3.28}
{
     {\tilde \omega}_\vq = {\tilde c} |\vq| 
     }
with $ {\tilde c} = \Omega_{\rm p}/{\tilde q}_s $. 
As is widely known, 
the longitudinal optical phonons are screened by electrons 
into the accoustic mode with a dispersion energy proportional to $|\vq|$ 
as expressed in \eq.{eq:accoustic}. 
It is easily verified that putting $U = 0$, 
one could recover 
all the standard results of the RPA of the electron gas model 
in the back-ground jellium positive charge~\cite{Sch83}. 
For $U \ne 0$, we find that 
the electron correlations due to the on-site Coulomb repulsion 
weaken the screening effect and increase the sound velocity.

%%% 4. 
\section{Superconductivity}

In this section, we examine the two-particle vertex part 
$\Gamma_{\sigma \sigma_1}(k,k')$, 
which induces superconductivity. 
We clarify the corrections to the vertex part 
due to the on-site Coulomb interactions.

%%% 4.1 
\subsection{Definition of the effective interactions}

The two-particle vertex part $\Gamma_{\sigma \sigma_1}(k,k')$ 
is defined as one which scatters 
the pair of two electrons with $(k,\sigma)$ and $(-k,\sigma_1)$ 
to that with $(k',\sigma)$ and $(-k',\sigma_1)$. 
The gap equation of superconductivity is written as 
\Equation{eq:gapeq}
{4.1}
{
     \begin{split}
     \Delta_{\sigma \sigma_1} (k)
     & 
     = - \sum_{k'} 
          \Gamma_{\sigma \sigma_1}(k,k') 
     \\[-8pt]
     & \hsp{8} 
           \times 
           G_{\sigma}(k') G_{\sigma_1}(-k')
           \Delta_{\sigma \sigma_1} (k') , 
     \end{split}
     }
where we have introduced an abbreviation 
\Equation{eq:sumabbreviation}
{4.2}
{
     \sum_{k'} = N^{-1} \sum_{\vk'} T \sum_{n'} 
     } 
and $G_{\sigma}(k)$ denotes the dressed electron Green's function. 
The vertex part $\Gamma_{\sigma \sigma_1}$ 
plays a role of the effective interactions. 
The gap equation \refeq{eq:gapeq} includes the dynamical effect 
by the Matsubara frequency.

The electron self-energy is calculated by 
\Equation{eq:selfenergy}
{4.3}
{
     \Sigma_{\sigma}(k) 
     = - \sum_{k' \sigma'} 
          \Gamma_{\sigma \sigma'}^{\rm (N)} (k,k') G_{\sigma'}(k') , 
     }
where $\Gamma_{\sigma \sigma'}^{\rm (N)}$ denotes 
the appropriate vertex part for the self-energy consistent with 
$\Gamma_{\sigma \sigma_1}$. 
The consistency is considered phenomenologically 
or microscopically with the Ward-Takahashi identity. 
Equations~\refeq{eq:gapeq} and~\refeq{eq:selfenergy} 
are called Eliashberg equations.

In order to estimate $\Tc$ accurately, 
we need to solve the full set of equations. 
Especially when we consider 
the spin-fluctuation-mediated superconductivity, 
since the attractive part of the interactions 
is much weaker than the repulsive part, 
the self-energy effect is important~\cite{NoteSEEonTc}. 
However, since in this paper we are mainly interested 
in the phonon-mediated pairing interactions, 
we omit the self-energy renormalization of the electrons 
from now on.

We can improve the perturbation theory~\cite{Bic89,Yon90} 
by including renormalization of the electron Green's function 
and the vertex corrections in $\Gamma_{\sigma,\sigma_1}(k,k')$. 
For example, we could take into account the correction 
to the spin and charge fluctuations 
from the self-energy effect by replacing $\chi_0(q)$ with $\chi(q)$, 
as in the renormalized RPA, 
although it requires heavy numerical calculations.

However, we expect that the most essential diagrams 
are taken into account in the RPA. 
Within the RPA, we could reproduce a physical situation that 
the strong charge and spin fluctuations mediate 
the pairing interactions. 
This consideration is partially based on a phenomelogical viewpoint. 
However, more or less, 
we could not avoid such a phenomenological consideration 
in the perturbation theories based on effective Hamiltonians, 
such as the Hubbard model. 
In any perturbation theory, we need to ignore almost all diagrams 
in the infinite series of the perturbation expansion. 
Taking a particular series of the diagrams can be rigorously justified 
only in an appropriate limit, 
but otherwise it is not quantitatively accurate very much. 
Furthermore, we need to use the effective Hamiltonian, 
which is usually far simpler than the real materials. 
Usually, effective Hamiltonians are useful for qualitative study, 
but are not very precise quantitatively. 
Therefore, even if we develope a perturbation theory 
by purely theoretical efforts beyond the RPA, 
the applicability of the quantitative results 
to the real materials are limited, 
unless any qualitatively distinct result is obtained.

%%% 4.2 
\subsection{Expression of the effective pairing interactions}

In this subsection, we derive the expression of 
$\Gamma_{\sigma \sigma_1}(k,k')$ from the results of 
the previous section \S 3. 
From the momentum and spin assignment depicted 
in Fig.~\ref{fig:Gammamomass} 
that 
$\Gamma^{\rm ring}_{\sigma \sigma'}(k,-k,k-k')$ and 
$\Gamma^{\rm lad}_{\sigma \sigma'}(k,-k,k-k')$ 
in eqs.~\refeq{eq:Vringelements} and \refeq{eq:Vladder} 
contribute to the vertex part $\Gamma_{\sigma \sigma_1}(k,k')$.

The contribution from the charge fluctuations and 
the screened Coulomb interaction is obtained from 
eqs.~\refeq{eq:Vringrewritten} and \refeq{eq:Vringelements} 
as 
\Equation{eq:Gammac}
{4.4}
{
     \Gamma_{\rm c}(q)
     = 
     \frac{1}{2} 
     \frac{2 v_{\vq} + U } 
          { 1 + (2 v_{\vq} + U) \chi_0(q) } 
     = \frac{v_{\vq} + U/2}{\kappa(q)} , 
     } 
where we have put $q = k - k'$. 
Similarly, the contribution from the spin fluctuations 
with respect to the $S_z$ component is written as 
\Equation{eq:Gammaszz}
{4.5}
{
     \sigma \sigma_1 
     \Gamma_{\rm s}^{zz}(q) 
     = 
     - 
     \frac{\sigma \sigma_1}{2} 
     \frac{U} 
          { 1 - U \chi_0(q) } . 
     }
In addition to them, 
we obtain the effective interaction mediated by phonons 
\Equation{eq:Gammaph}
{4.6}
{
     \Gamma_{\rm ph}(q) 
          = [{\tilde M}(q)]^2 \mathcal{D}(q) 
     = 
         \frac{M_{\vq}^2}{\kappa(q)} \cdot 
         \frac{- 2 \Omega_{\rm p}/\kappa(q)} 
              {\nu_m^2 + [{\tilde \omega}(q)]^2 } 
     }
with $\tilde \omega(q)$ given in \eq.{eq:omegatilde}.

The contribution from the spin fluctuations 
with respect to the $S_x$ and $S_y$ components is obtained 
from the ladder diagram as 
\Equation{eq:Gammasxy}
{4.7}
{
     \begin{split}
     {\bar \delta}_{\sigma \sigma_1}
     \Gamma_{\rm s}^{+-}(k,k') 
          & = 
            {\bar \delta}_{\sigma \sigma_1}
            \Gamma^{\rm lad}(k,-k,k-k') 
     \\
          & = 
          {\bar \delta}_{\sigma \sigma_1}
          \frac{U}{1 - U \chi_0(k + k')} , 
     \end{split}
     }
where ${\bar \delta}_{\sigma \sigma_1} = 1 - \delta_{\sigma \sigma_1}$. 
By the parity of the gap function $\Delta(\vk,{\rm i} \omega_{n})$, 
we could replace the vertex function \eq.{eq:Gammasxy} 
in the gap equation \eq.{eq:gapeq} with 
\Equation{eq:Gammasxyreplacedmom}
{4.8}
{
     {\bar \delta}_{\sigma \sigma_1}
     \Gamma_{\rm s}^{+-}(k,k') 
     = 
        s {\bar \delta}_{\sigma \sigma_1} 
          \frac{U}{1 - U \chi_0(k - k')} , 
     }
where we take the sign $s = + 1$ and $s = - 1$ 
for singlet pairing and triplet pairing, respectively.

Collecting the contributions of 
{eqs.~\refeq{eq:Gammac} - \refeq{eq:Gammasxyreplacedmom}}, 
we obtain the vertex part 
\Equation{eq:Gammaeff}
{4.9}
{
     \begin{split}
     \Gamma_{\sigma \sigma_1}(k,k') 
     & = 
         \Gamma_{\rm c}(k-k') 
       + 
         \sigma \sigma_1
         \Gamma_{\rm s}^{zz}(k-k') 
     \\ 
     & \hsp{4} 
       + \Gamma_{\rm ph}(k-k') 
       + {\bar \delta}_{\sigma \sigma_1} 
         \Gamma_{\rm s}^{+-}(k,k') 
     \\ 
     & \hsp{4} 
       - U {\bar \delta}_{\sigma \sigma_1} . 
     \end{split}
     }
Here, the last term $ - U {\bar \delta}_{\sigma \sigma_1}$ 
is to subtract the double counted contribution 
as mentioned below \eq.{eq:HCeff}. 
With a definition 
\Equation{eq:chispin}
{4.10}
{
     \Gamma_{\rm s}(q) 
          \equiv \frac{1}{2} \frac{U}{1 - U \chi_0(q)} , 
     }
\eq.{eq:Gammaeff} can be rewriten as 
\Equation{eq:Gammaeff2}
{4.11}
{
     \Gamma_{\sigma \sigma_1} 
     = 
         \Gamma_{\rm c}
       - \sigma \sigma_1 \Gamma_{\rm s} 
       + \Gamma_{\rm ph} 
       + 2 {\bar \delta}_{\sigma \sigma_1} 
         s \Gamma_{\rm s} 
       - U {\bar \delta}_{\sigma \sigma_1} . 
     }
Therefore, for singlet pairing, 
the vertex part $\Gamma_{\sigma,\sigma_1}(k,k')$ 
is expressed as 
\Equation{eq:Gammaeffsinglet}
{4.12}
{
     \Gamma_{\rm sp} 
     = 
         \Gamma_{\rm c}
       + 3 \Gamma_{\rm s} 
       + \Gamma_{\rm ph} 
       - U , 
     }
while 
for triplet pairing 
\Equation{eq:Gammaefftriplet}
{4.13}
{
     \Gamma_{\rm tp} 
     = 
         \Gamma_{\rm c}
       + \Gamma_{\rm s} 
       + \Gamma_{\rm ph} . 
     }
If we put $v_{\vq} = 0$, 
the spin and charge fluctuation terms are reduced to 
the expressions obtained 
in many articles~\cite{Sca86,NoteCEX,Shi88,Shi89,Shi00e}. 
Due to the complicated momentum dependence, 
it can work as an attractive interaction or as a repulsive interaction 
in the formation of the Cooper pairs, 
depending on the symmetry of the gap function 
and the shape of the Fermi surface.

Since we have obtained the expressions of the vertex part 
eqs.~\refeq{eq:Gammaeffsinglet} and \refeq{eq:Gammaefftriplet}, 
it is possible in principle to solve the Eliashberg equations 
\refeq{eq:gapeq} and \refeq{eq:selfenergy}. 
However, we leave it for future study, 
and derive weak coupling effective Hamiltonian 
in the next section \S5.

%%% 4.3 
\subsection{Physical interpretation}

In this subsection, we rewrite the expressions obtained above 
for a physical interpretation. 
We also discuss the expressions in the two limits, 
$U \rightarrow 0$ and $v_{\vq} \rightarrow 0$.

In \eq.{eq:Gammaeff2}, 
the screened Coulomb interactions and 
the phonon-mediated interactions are rewritten as 
\Equation{eq:Gammac_plus_Gammaph}
{4.14}
{
     \begin{split}
     \Gamma_{\rm c} + \Gamma_{\rm ph} 
     & = 
     \frac{v_{\vq} + U/2}{\kappa } 
     - 
     \frac{v_{\vq}}{\kappa \kappa_{U}} 
         \frac{[{\tilde \omega}(q)]^2} 
              {-\omega^2 + [{\tilde \omega}(q)]^2 } 
     \\
     & = 
       \frac{U}{2 \kappa_U } 
     + \frac{v_{\vq}}{\kappa \kappa_{U}} 
         \frac{ \omega^2 } 
              { \omega^2 - [{\tilde \omega}(q)]^2 } , 
     \end{split}
     }
where we have made the analytic continuation 
${\rm i}\nu_m \rightarrow \omega \pm {\rm i}\delta$, 
and put $q = k - k'$.

Therefore, the total two-particle vertex part is given by 
\Equation{eq:Gammaefftheta}
{4.15}
{
     \Gamma_{\sigma \sigma_1}(q) 
     = \Gamma_U^{(0)} 
     + \frac{v_{\vq}}{\kappa \kappa_{U}} 
         \frac{ \omega^2 } 
              { \omega^2 - [{\tilde \omega}(q)]^2 } , 
     }
with $q = k - k'$, 
where the function $\Gamma_U^{(0)}$ in the first term 
is nothing but the vertex part in the pure Hubbard model, 
{\it i.e.}, 
the model with $U \ne 0$ and $v_{\vq} = 0$~\cite{Sca86,Shi88,Shi89}. 
For singlet pairing, it is expressed as 
\Equation{eq:GammaU0singlet}
{4.16}
{
     \begin{split}
     \Gamma_{U}^{(0)} 
     & = 
         \frac{1}{2} \frac{U}{1 + U \chi_0}
       + \frac{3}{2} \frac{U}{1 - U \chi_0}
       - U 
     \\ 
     & = 
         \frac{U}{1 - U^2 \chi_0^2 }
       + \frac{U}{1 - U \chi_0}
       - U , 
     \\
     \end{split}
     }
which can be written as 
\Equation{eq:GammaU0singletHubbardsus}
{4.17}
{
     \Gamma_{U}^{(0)} 
     = 
       - \frac{1}{4} U^2 \chi_{\rm c}^{(0)} 
       + 3 U^2 \chi_{\rm s}^{(0)} 
       + U 
     }
with \eq.{eq:susceptibilityHubbard}. 
For triplet pairing, it is expressed as 
\Equation{eq:GammaU0triplet}
{4.18}
{
     \Gamma_{U}^{(0)} 
     = 
         \frac{1}{2} \frac{U}{1 + U \chi_0}
       + \frac{1}{2} \frac{U}{1 - U \chi_0}
     = \frac{U}{1 - U^2 \chi_0^2} , 
     }
which can be written as 
\Equation{eq:GammaU0tripletHubbardsus}
{4.19}
{
     \Gamma_{U}^{(0)} 
     = 
       - \frac{1}{4} U^2 \chi_{\rm c}^{(0)} 
       + U^2 \chi_{\rm s}^{(0)} 
       + U . 
     }
In eqs.~\refeq{eq:GammaU0singletHubbardsus} 
and~\refeq{eq:GammaU0tripletHubbardsus}, 
the physical interpretation of each term is 
clear~\cite{Sca86,Shi88,Shi89}. 
The first terms of eqs.~\refeq{eq:GammaU0singletHubbardsus} 
and~\refeq{eq:GammaU0tripletHubbardsus} 
are the interactions mediated by the charge fluctuations, 
while the second terms are those by the spin fluctuations. 
Obviously, the last constant terms in 
eqs.~\refeq{eq:GammaU0singletHubbardsus} 
 and~\refeq{eq:GammaU0tripletHubbardsus} 
express the bare on-site Coulomb repulsive interactions, 
and do not contribute to the gap equation 
for anisotropic pairing. 
If we put $v_{\vq} = 0$, we obtain 
\Equation{eq:Gammav0}
{4.20}
{
     \Gamma_{\sigma \sigma_1} = \Gamma_U^{(0)} , 
     }
and all the equations are reduced to those in the Hubbard model.

The second term of \eq.{eq:Gammaefftheta} corresponds to 
the summation of 
the screened Coulomb and phonon-mediated interactions. 
If we put $U = 0$, we obtain 
\Equation{eq:GammaU0}
{4.21}
{
     \Gamma_{\sigma \sigma_1}(q) 
     = 
       \frac{v_{\vq}}{\kappa_v} 
         \frac{ \omega^2 } 
              { \omega^2 - [\omega(q)]^2 } 
     \equiv 
     \Gamma (q) , 
     }
where $\kappa_v = 1 + 2 v_{\vq} \chi_0$ and 
$\omega(q) = \Omega_{\rm p}/\sqrt{\kappa_v(q)}$. 
This equation is easily rewritten as 
\Equation{eq:Gammaeff2U0}
{4.22}
{
     \begin{split}
     \Gamma_{\sigma \sigma_1}(q) 
     & = 
         \frac{v_{\vq}}{\kappa_v} 
       + [{\tilde M}(q)]^2 \mathcal{D}(q) 
     \\ 
     & = 
         \frac{v_{\vq}}{\kappa_v} 
       + \frac{ M_{\vq}^2 }{ \kappa_v^2 }
         \frac{ 2 \Omega_{\rm p} }
              { \omega^2 - [\omega(q)]^2 } . 
     \end{split}
     }
We note that the expression of $\Gamma_{\sigma \sigma_1}$ 
in the presence of both $U$ and $v_{\vq}$, 
{\it i.e.}, \eq.{eq:Gammaefftheta}, 
is not obtained by simply adding 
eqs.~\refeq{eq:Gammav0} and~\refeq{eq:GammaU0}. 
The second term of \eq.{eq:Gammaefftheta} includes 
the corrections due to $U$ in a rather complicated manner 
as explained in the next subsection.

%%% 4.4 
\subsection{Interpretation of the corrections due to $U$}

Now, let us interpret how the corrections from $U$ enters 
in the phonon-mediated pairing interaction, {\it i.e.}, 
the second term of \eq.{eq:Gammaefftheta}. 
We can derive \eq.{eq:Gammaefftheta} in another manner as follows, 
starting from \eq.{eq:Gammaeff2U0}, {\it i.e.}, 
the effective interaction in the absence of $U$.

First, we rewrite \eq.{eq:Gammaeff2U0} as 
\Equation{eq:Gammaeff2U0rewrite}
{4.23}
{
     \Gamma_{\sigma \sigma_1}(q) 
     = 
         \frac{v_{\vq}}{\kappa_v} 
       + \frac{ 2 M_{\vq}^2 }{ \kappa_v \Omega_{\rm p} }
         \frac{ [\omega(q)]^2 }
              { \omega^2 - [\omega(q)]^2 } , 
     }
and in this equation we replace $\kappa_v$ 
with ${\tilde \kappa}_v$ renormalized 
by the Hubbard-type charge fluctuations 
as shown in \eq.{eq:kappavtilde}. 
At the same time, we replace the phonon frequency $\omega(q)$ 
with ${\tilde \omega}(q)$ of the fully dressed phonons, 
which is given by \eq.{eq:omegatilde}. 
Secondly, we modify the electron-phonon coupling constant $M_{\vq}$ 
with 
$$
     \frac{M_{\vq}}{\kappa_U} 
     = 
     M_{\vq}
          \left [ 
               1 - U \chi_0 + U^2 \chi_0^2 \mp \cdots 
          \right ] . 
     $$
The series in the right-hand-side is easily verified 
from the diagramatical technique. 
At the same time, we insert $1/\kappa_U$ into the both end of 
the screened Coulomb line which corresponds to 
$v_{\vq}/{\tilde \kappa_v}(q)$. 
In other words, we replace $v_{\vq}/{\tilde \kappa_v}$ 
with $v_{\vq}/{\tilde \kappa_v}\kappa_U^2$. 
Thirdly, 
we add the Hubbard-type fluctuation diagrams $\Gamma_U^{(0)}$ 
including bare $U$. 
Then, we obtain the expression of $\Gamma_{\sigma \sigma_1}(q)$ 
\Equation{eq:Gammaeff2U0rewrite2}
{4.24}
{
     \Gamma_{\sigma \sigma_1}(q) 
     = 
     \Gamma_U^{(0)} 
       + \frac{v_{\vq}}{ \kappa_U^2 {\tilde \kappa}_v} 
       + \frac{ 2 M_{\vq}^2 }{ \kappa_U^2 \kappa_v \Omega_{\rm p} }
         \frac{ [{\tilde \omega}(q)]^2 }
              { \omega^2 - [{\tilde \omega}(q)]^2 } , 
     }
which coincides with \eq.{eq:Gammaefftheta}, 
since 
\Equation{eq:vertexrenormalizations}
{4.25}
{
     \frac{v_{\vq}}{\kappa \kappa_{U}} 
     = 
     \frac{1}{\kappa_U^2} 
     \frac{v_{\vq}}{{\tilde \kappa}_v} 
     = 
     \frac{2 M_{\vq}^2}{\kappa_U^2 {\tilde \kappa}_v \Omega_{\rm p}} 
     }
holds from eqs.~\refeq{eq:diele2} and \refeq{eq:ephcoupling}.

\section{Effective Interactions for Weak Coupling}

In this section, 
we derive the effective pairing interactions 
along the scheme of the traditional weak coupling theory. 
The effective interactions derived here are not suitable for 
accurate estimations for the transition temperature, 
but useful for qualitative and semi-quantitative arguments. 
For quantitative purpose, 
direct numerical calculations based on 
the effective vertex eqs.~\refeq{eq:Gammaeffsinglet} 
and \refeq{eq:Gammaefftriplet} is more suitable.

%%% 5.1 
\subsection{The case of \hsp{0} $U = 0$}

First, we briefly review the standard derivation of 
the effective pairing interactions for $U = 0$~\cite{Sch83}. 
In this case, the total two-particle vertex part 
$\Gamma_{\sigma \sigma_1}$ 
is given by \eq.{eq:GammaU0}. 
In this equation, it is explicitly shown that 
the phonon-mediated interaction is attractive for small $\omega$ 
overcoming repulsive Coulomb interactions. 
This effect is called the overscreening effect.

\begin{figure}
%% JPSJ %% When you do not use epsf, activate the next line. 
%% \figureheight{7cm}
%% 
%% PR %% \begin{figure}[htb]
\begin{center}
%% FOR TWO COLUMN 
%% \leavevmode \epsfxsize=6cm  
%% 
%% FOR ONE COLUMN 
%% \leavevmode \epsfxsize=10cm  
%% JPSJ2 %% 
\vspace{2\baselineskip}
\includegraphics[width=6.5cm]{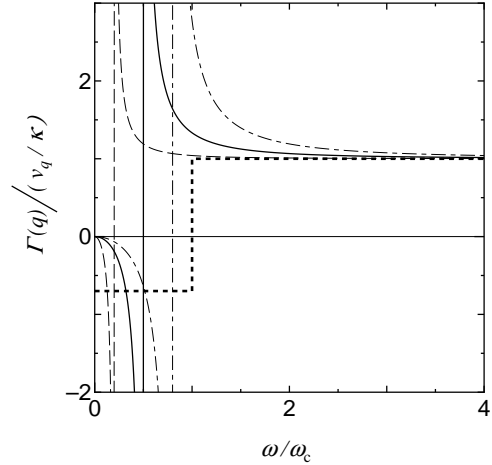}
\vspace{\baselineskip}
%% PR %% \leavevmode \epsfxsize=7cm  
%% PR %% \epsfbox{fig08.eps}
\end{center}
\caption{
Schematic figure of the two-particle vertex part 
$\Gamma(q)$ as a function of $\omega/\omega_{\rm c}$. 
The solid, dashed, dot-dashed lines show $\Gamma(q)$ 
as functions of $\omega$ with different fixed values of $q$. 
The thick dotted line shows a step function which is introduced to 
approximate $\Gamma(q)$ in \eq.{eq:Gammastepfunction}. 
} 
\label{fig:Gamma_approx_by_thetafunc}
\end{figure}

In the weak coupling theory, 
the frequency dependence of the two-particle vertex part 
$\Gamma(q)$ is simplified by a replacement 
with a step function as shown 
in Fig.~\ref{fig:Gamma_approx_by_thetafunc}, 
that is, 
\Equation{eq:Gammastepfunction}
{5.1}
{
     \Gamma(\vq,\omega) 
     \approx 
         \frac{v_{\vq}}{\kappa(\vq,0)} 
         \left [ 
           1 - {\bar \alpha} \, \theta ( \omega_{\rm c} - |\omega| ) 
         \right ] , 
     }
where $\omega_{\rm c}$ and ${\bar \alpha}$ denote effective constants 
which express the frequency range and 
the depth of the attractive part of the interactions. 
Here, the static approximation for the dielectric function 
has been introduced.

The effective constants ${\bar \alpha}$ and $\omega_{\rm c}$ 
are complicated quantities that reflect the dispersion energy 
and the density of states of the dressed phonons. 
However, it is reasonable to consider that 
the constant $\omega_{\rm c}$ is of the order of 
the Debye frequency $\omega_{\rm D}$, {\it i.e.}, 
the upper limit of the dressed (observed) phonon frequency 
${\tilde \omega}_{\vq}$. 
According to the standard weak coupling theory, 
we regard those constants as parameters in this paper.

For isotropic superconductors, the prefactor $v_{\vq}/\kappa(\vq,0)$ 
is often replaced by an effective constant $V_0$. 
We should note that $V_0$ is an averaged quantity of $v_{\vq}$ 
and different from the on-site Coulomb interaction 
$U$ derived in \S 2~\cite{Shi02a,Shi02b}.

Further, in the weak coupling theory, 
the cutoff $\omega_{\rm c}$ in $\omega$ integral is taken into account 
by introducing the same cutoff in $\xi$ integral, 
where $\xi$ denotes the electron energy 
measured from the Fermi energy. 
In other words, 
we replace the cutoff function 
$\theta ( \omega_{\rm c} - |\omega| )$ 
with 
$ \theta (\omega_{\rm c} - |\xi_{\vk}|) 
  \theta (\omega_{\rm c} - |\xi_{\vk'}|) $.

After these simplifications, 
we could solve the gap equation 
along the scheme of the standard weak coupling theory. 
The Coulomb repulsion, {\it i.e.}, the first term in the braket 
of \eq.{eq:Gammastepfunction}, is taken into account 
through the effective Coulomb parameter defined by 
\Equation{eq:effectiveCoulomb0}
{5.2}
{
     \mu_{V0}^{*}
     = 
     \frac{V_0 N(0)}{ 1  +  V_0 N(0) \ln (W_{\rm C}/\omega_{\rm c}) } , 
     }
where $W_{\rm C}$ denotes the cutoff energy of the Coulomb interactions, 
which is of the order of the band width $W$~\cite{Notemuast}. 
The influence of the Coulomb repulsion is weaken 
($\mu_{V0}^{*} < V_0 N(0)$) 
by the retardation of the phonon propagation 
when $\omega_{\rm c} < W_{\rm C}$. 
The superconducting transition temperature $\Tc$ is given by 
\Equation{eq:BCS}
{5.3}
{
     \Tc = 1.13 \, \omega_{\rm c} \e^{- 1/(\lambda - \mu_{V0}^{*})} , 
     }
where $\lambda = {\bar \alpha} V_0 N(0)$.

%%%%%%%%%%%%%%%%%%%%%%%%%%%%%%%%%%%%%%%%%%%%%%%%%%%%%%%%%%%%%%%%%%%%%%%
%%  Fig.8                                                            %%
%%%%%%%%%%%%%%%%%%%%%%%%%%%%%%%%%%%%%%%%%%%%%%%%%%%%%%%%%%%%%%%%%%%%%%%
%% JPSJ %% 
%%%%%%%%%%%%%%%%%%%%%%%%%%%%%%%%%%%%%%%%%%%%%%%%%%%%%%%%%%%%%%%%%%%%%%%

%%% 5.2 
\subsection{The case of \hsp{0} $U \ne 0$}

We extend the above formulation to the case of $U \ne 0$. 
We introduce the same simplification 
as \eq.{eq:Gammastepfunction} 
in the context of the weak coupling theory. 
Hence, it is obtained that 
\Equation{eq:Gammac_plus_Gammaph_in_theta}
{5.4}
{
     \Gamma_{\sigma \sigma_1}(\vq,\omega) 
     = \Gamma_U^{(0)} 
     + {\tilde v}_{\vq} 
         \left [ 
           1 - {\bar \alpha} \, \theta ( \omega_{\rm c} - |\omega| ) 
         \right ] , 
     }
with ${
     {\tilde v}_{\vq} 
     \equiv {v_{\vq}}/{\kappa(\vq,0) \kappa_{U}(\vq,0)}  
     }$, 
where the static approximation has been also introduced 
in $\kappa$, $\kappa_U$ and $\Gamma_U^{(0)}$ 
as in \eq.{eq:Gammastepfunction}.

Let us compare the expressions for $U = 0$ and $U \ne 0$, 
{\it i.e.}, eqs.~\refeq{eq:Gammastepfunction} 
and~\refeq{eq:Gammac_plus_Gammaph_in_theta}. 
One of the important differences is that 
\eq.{eq:Gammac_plus_Gammaph_in_theta} includes 
the Hubbard-type effective interaction $\Gamma_U^{(0)}$ 
that we have discussed in \S 4.3.

Another difference between eqs.~\refeq{eq:Gammastepfunction} 
and~\refeq{eq:Gammac_plus_Gammaph_in_theta} is 
the renormalization 
of the second term of \eq.{eq:Gammac_plus_Gammaph_in_theta} 
due to the on-site Coulomb interaction $U$. 
In this term, the prefactor ${\tilde v}_{\vq}$ 
and the constants ${\bar \alpha}$ and $\omega_{\rm c}$ are modified. 
The constants ${\bar \alpha}$ and $\omega_{\rm c}$ are 
usually treated as parameters in the weak coupling theory, 
and sometimes determined by experiments. 
Hence, 
we do not need to know the differences from their original values, 
and the modifications in ${\bar \alpha}$ and $\omega_{\rm c}$ 
do not need to be considered explicitly. 
Therefore, the important modification is 
in the prefactor ${\tilde v}_{\vq}$, 
because it affects the momentum dependence. 
From eqs.~\refeq{eq:diele2} and \refeq{eq:kappavtilde}, 
it is written as 
\Equation{eq:PhMInt_factor}
{5.5}
{
     {\tilde v}_{\vq}
     \equiv 
     \frac{v_{\vq}}{\kappa \kappa_U} 
     = \frac{1}{2 \chi_0 \kappa_U } 
     \frac{{\tilde q}_{s}^2 }{ |\vq|^2 + {\tilde q}_{s}^2 }  . 
     }
Putting $U = 0$ in \eq.{eq:PhMInt_factor}, 
it reduces to the factor of \eq.{eq:Gammastepfunction}, 
{\it i.e.}, 
\Equation{eq:PhMInt_factor_U0}
{5.6}
{
     \frac{1}{2 \chi_0} 
     \frac{q_{\rm s}^2 }{ |\vq|^2 + q_{\rm s}^2 } . 
     }
Comparing eqs.~\refeq{eq:PhMInt_factor} and~\refeq{eq:PhMInt_factor_U0}, 
we note that $q_{\rm s}$ is replaced 
with a smaller quantity ${\tilde q}_{s}$ 
in \eq.{eq:PhMInt_factor}. 
As discussed in \S 3.2, 
the screening length 
$\lambda_{\rm scr} = q_{\rm TF}^{-1} = [q_{\rm s}]_{q = 0}^{-1}$ 
is lengthened into 
${\tilde \lambda}_{\rm scr} = [{\tilde q}_{\rm s}]_{q = 0}^{-1}$ 
because $U > 0$. 
Then, the peak of \eq.{eq:PhMInt_factor} becomes sharper, 
and the anisotropic components increase.

In summary of this part, we find two effects of the short-range 
correlations: 
(1)~The Hubbard term $\Gamma_U^{(0)}$, which includes bare $U$, 
is added as pointed out in the previous section \S4; 
(2)~The short-range correlations change the momentum dependence 
of the phonon-mediated pairing interaction. 
Both effects unfavor isotropic pairing, 
but favor anisotropic pairing of some symmetries.

It is also found that for this mechanism of triplet pairing, 
the existence of the on-site Coulomb interaction $U$ is essential, 
but not the $s$-wave (isotropic) component of $v_{\vq}$, 
such as $V_0$. 
We note in the second term of 
\eq.{eq:Gammac_plus_Gammaph_in_theta} 
that the long-range part ${\tilde v}_{\vq}$ of the Coulomb interaction 
and the phonon-mediated pairing interaction 
$- \alpha {\tilde v}_{\vq} \theta(\omega_{\rm c} - |\omega|)$ 
have the same momentum dependence. 
Hence, it is not effective for the change in the dominancy of 
the coupling constants. 
In fact, when we put the dominant isotropic component $V_0$ 
and the subdominant anisotropic component $V_1$, 
the difference of the effective coupling constants 
which include the negative contributions of 
the effective Coulomb parameters is calculated as 
\Equation{eq:dominancy_onlylongrangepart}
{5.7}
{
     \begin{array}{l}
     \dps{ 
     \Bigl [
     {\bar \alpha} V_0
     - \frac{V_0}{1 + V_0 x}
     \Bigr ]
     - 
     \Bigl [
     {\bar \alpha} V_1 
     - \frac{V_1}{1 + V_1 x}
     \Bigr ]
     }
     \\[8pt]
     \dps{ 
     = (V_0 - V_1) 
          \Bigl [
          {\bar \alpha} 
               - \frac{1}{(1 + V_0 x)(1 + V_1 x)}
          \Bigr ] , 
     }
     \end{array}
     }
where $x \equiv N(0) \ln(W_{\rm C}/\omega_{\rm c})$. 
We have ${\bar \alpha} > 1$ when the overscreening effect gives rise to 
a net attractive interaction for low frequencies. 
Thus, when $V_0 > V_1$, 
the right-hand-side of \eq.{eq:dominancy_onlylongrangepart} 
is positive. 
Therefore, it is found that 
the long-range part ${\tilde v}_{\vq}$, which gives rise to 
the third and fourth terms in the left-hand-side of 
\eq.{eq:dominancy_onlylongrangepart}, 
does not change the sign of \eq.{eq:dominancy_onlylongrangepart}. 
In the rest of this paper, 
we investigate the change 
in the dominancy of the coupling constants 
due to the Coulomb interaction more explicitly.

%%% 5.3 
\subsection{Interactions mediated by spin fluctuations}

In this subsection, we develope a weak coupling theory 
including interactions mediated by spin fluctuations 
through $\Gamma_U^{(0)}$. 
In particular, when the spin fluctuations are relevant, 
the vertex part $\Gamma_U^{(0)}(\vq,\omega)$ 
consists of two parts as a function of $\vq$ and $\omega$: 
(a) A broad peak that reflects the energy scale of 
the band width $\sim W$; 
(b) A sharp peak that reflects the energy scale of 
the spin fluctuations $\omega_{\rm sf}$, 
as discussed in our previous papers~\cite{Shi03b,Shi89}. 
Hence, we write it in the form 
\Equation{eq:GammaUtwoparts}
{5.8}
{
     \Gamma_U^{(0)}(\vq,\omega) 
     \approx 
     {\bar \Gamma}_U^{(0)}(\vq,0) 
     + 
     \Gamma_{\rm sf}^{(0)}(\vq,0) 
     \theta (\omega_{\rm sf} - |\omega|) , 
     }
where the first and second terms which correspond to (a) and (b) 
mentioned above, respectively. 
Obviously, 
${\bar \Gamma}_U^{(0)}$ includes the on-site Coulomb repulsive 
interaction. 
Therefore, we obtain 
\Equation{eq:Gammatwocutoffomega}
{5.9}
{
     \begin{split}
     \Gamma_{\sigma \sigma_1}(\vq,\omega) 
     & = 
     V_{\rm C}(\vq) 
     + \Gamma_{\rm sf}^{(0)}(\vq,0) 
         \theta ( \omega_{\rm sf} - |\omega| ) 
     \\
     & \hsp{4} 
     - {\bar \alpha} {\tilde v}_{\vq} \, 
         \theta ( \omega_{\rm c} - |\omega| ) 
     \end{split}
     }
from \eq.{eq:Gammac_plus_Gammaph_in_theta}, 
where we have defined a Coulomb term 
$V_{\rm C}(\vq) \equiv {\bar \Gamma}_U^{(0)}(\vq,0)
+ {\tilde v}_{\vq}$.

As mentioned in \S 5.1, 
we replace 
$\theta ( \omega_k - |\omega| )$ 
with 
$ \theta (\omega_k - |\xi_{\vk}|) 
  \theta (\omega_k - |\xi_{\vk'}|) $ 
within the weak coupling theory, 
where $\omega_1 = \omega_{\rm c}$ and $\omega_2 = \omega_{\rm sf}$. 
Therefore, we obtain effective interactions 
\Equation{eq:Veffwithsfandph}
{5.10}
{
     V_{\rm eff}(\vk,\vk') 
       = V_{\rm C}(\vk - \vk') 
         + V_{\rm sf}(\vk,\vk') 
         + V_{\rm ph}(\vk,\vk') , 
     }
where 
\Equation{eq:VsfeffVpheff}
{5.11}
{
     \begin{split}
     V_{\rm sf}(\vk,\vk') 
     & = \Gamma_{\rm sf}^{(0)}(\vk - \vk',0)
         \theta (\omega_{\rm sf} - |\xi_{\vk}|) 
              \theta (\omega_{\rm sf} - |\xi_{\vk'}|) 
     \\
     V_{\rm ph}(\vk,\vk') 
     & = - {\bar \alpha} {\tilde v}_{\vk - \vk'} 
              \theta (\omega_{\rm c} - |\xi_{\vk}|) 
              \theta (\omega_{\rm c} - |\xi_{\vk'}|) . 
     \end{split}
     }
When $\omega_{\rm sf} \ll W$ and $\omega_{\rm c} \ll W$, 
we may regard $\Gamma_{\rm sf}^{(0)}$ and ${\tilde v}_{\vk - \vk'}$ 
as functions of $k_{\parallel}$ and $k'_{\parallel}$, 
which denote the momentum components of $\vk$ and $\vk'$ 
along the Fermi surface. 
Hence, we introduce a complete set of the basis functions 
$\gamma_{\alpha}(k_{\parallel})$ with $\alpha = 0,1,2, \cdots$, 
which are normalized by the average on the Fermi surface. 
For example, we may replace $k_{\parallel}$ with 
${\hat \vk} \equiv \vk/|\vk|$ or $(\theta,\varphi)$, 
and take $\alpha = s, p_x, p_y, p_x \cdots $ or $\alpha = (l,m)$ 
in the spherically symmetric systems. 
We define 
$V_{\rm sf}(k_{\parallel},k'_{\parallel}) 
     \equiv [\Gamma_{\rm sf}^{(0)}(\vk - \vk',0)]_{\xi_\vk = \xi_\vk'= 0}$ 
and 
$V_{\rm ph}(k_{\parallel},k'_{\parallel}) 
     \equiv - {\bar \alpha} [{\tilde v}_{\vk - \vk'}]_{\xi_\vk = \xi_\vk'= 0}$, 
and expand the interactions as 
\Equation{eq:Veffexpand}
{5.12}
{
     V_k(k_{\parallel},k'_{\parallel}) 
     = \sum_{\alpha} V_k^{\alpha} 
          \gamma_{\alpha}(k_{\parallel}) 
          \gamma_{\alpha}(k'_{\parallel}) , 
     }
where the index $k$ donotes the kind of interactions, 
for example, $V_1 = V_{\rm ph}$, $V_2 = V_{\rm sf}$ 
and $V_3 = V_{\rm C}$. 
We define dimensionless coupling constants 
$\lambda_{\rm ph}^{\alpha} = - V_{\rm ph}^{\alpha} N(0)$ and 
$\lambda_{\rm sf}^{\alpha} = - V_{\rm sf}^{\alpha} N(0)$, 
and effective Coulomb parameter 
$\mu_{\alpha}^{*} 
     = V_{\rm C}^{\alpha} N(0)
       /[ 1 + V_{\rm C}^{\alpha} N(0) \ln(W_{\rm C}/\omega_{\rm c})]$.

The gap function is also expanded as 
\Equation{eq:Deltaalphadef}
{5.13}
{
     \Delta (\vk) 
     = 
     \sum_{\alpha} \Delta_{\alpha} 
          \gamma_{\alpha}(k_{\parallel})
     } 
on the Fermi surface. 
By taking the basis functions $\gamma_{\alpha}$ appropriately, 
we can make the gap equation diagonal with respect to 
the symmetry index $\alpha$ upto the linear order of $\Delta$. 
While each decoupled gap equation gives 
a different second order transition temperature $T_{{\rm c}\alpha}$, 
we must take the highest one of $T_{{\rm c}\alpha}$'s, 
since the system exhibits superconductivity 
when at least one of $\Delta_{\alpha}$'s is nonzero.

In the same way as in our previous paper~\cite{Shi03b}, 
we can obtain the transition temperature $\Tc$ 
and the isotope effect coefficient $\alpha_{\rm IE}$ 
defined by 
\Equation{eq:ie_alpha_def}
{5.14}
{
     \alpha_{\rm IE} = - \pardif{\ln \Tc}{\ln M} , 
     }
where $M$ denotes the relevant atomic mass. 
Because the application is straightforward, 
we shall omit the derivation and only present the results.

When $\omega_{\rm sf} < \omega_{\rm c} \sim \omega_{\rm D}$, 
which occurs in the proximity of the magnetic instability, 
we obtain the transition temperature 
$\Tc = 1.13 \, \omega_{\rm c} \exp[-1/(   \lambda_{\rm sf}^{\alpha} 
                                     + \lambda_{\rm ph}^{\alpha *}) ]$ 
with 
$\lambda_{\rm ph}^{\alpha *} = (\lambda_{\rm ph}^{\alpha} - \mu^{*}_{\alpha})
 /[1 - (\lambda_{\rm ph}^{\alpha}- \mu^{*}_{\alpha}) 
        \ln (\omega_{\rm c}/\omega_{\rm sf})]$ 
and the isotope effect coefficient 
\Equation{eq:Tcalpha}
{5.15}
{
     \alpha_{\rm IE} 
       = \frac{1}{2} 
          \Bigl ( \frac{\lambda_{\rm ph}^{\alpha *}}
                       {\lambda_{\rm sf}^{\alpha} 
                         + \lambda_{\rm ph}^{\alpha *}}
          \Bigr )^2 
          \Bigl [ 
               1 - 
            \Bigl ( \frac{\mu^{*}_{\alpha}}
                    {\lambda_{\rm ph}^{\alpha} - \mu^{*}_{\alpha}}
            \Bigr )^2 
          \Bigr ] . 
     }
When $\omega_{\rm sf} > \omega_{\rm c} \sim \omega_{\rm D}$, 
which occurs away from the magnetic instability, 
we obtain the transition temperature 
$\Tc = 1.13 \, \omega_{\rm c} \exp[-1/(   \lambda_{\rm ph}^{\alpha} 
                                     + \lambda_{\rm sf}^{\alpha *}) ]$ 
with 
$\lambda_{\rm sf}^{\alpha *} 
 = (\lambda_{\rm sf}^{\alpha} - \mu_{\rm sf}^{\alpha *})
   /[1 - (\lambda_{\rm sf}^{\alpha} - \mu_{\rm sf}^{\alpha *}) 
          \ln (\omega_{\rm sf}/\omega_{\rm c})]$ 
and $\mu_{\rm sf}^{\alpha *} 
     = V_{\rm C}^{\alpha} N(0)
       /[ 1 + V_{\rm C}^{\alpha} N(0) \ln(W_{\rm C}/\omega_{\rm sf})]$. 
The isotope effect coefficient is obtained as 
\Equation{eq:alphaIE}
{5.16}
{
     \alpha_{\rm IE} 
       = \frac{1}{2} 
          \Bigl [ 
               1 - 
            \Bigl ( \frac{\lambda_{\rm sf}^{\alpha *}}
                       {   \lambda_{\rm ph}^{\alpha} 
                         + \lambda_{\rm sf}^{\alpha *}}
            \Bigr )^2 
          \Bigr ] . 
     }

%%% 5.4 
\subsection{Anisotropic pairing mediated by phonons}

In this subsection, we derive expressions of 
the effective coupling constants and the effective Coulomb parameters 
for anisotropic pairing. 
We concentrate our attention on the interactions mediated by phonons, 
and ignore the attractive interactions by $\Gamma_U^{(0)}$.

Hence, we consider the case in which we could ignore 
the momentum and frequency dependences of $\chi_0(q)$, 
so that $\chi_0(q) \approx \chi_0({\vector 0},0) = N(0)$ holds. 
Although this simplification becomes qualitatively better 
away from the magnetic instability, 
even there the spin fluctuations still have an important effect 
of enhancement of the repulsive Coulomb interactions as 
\Equation{eq:spinflucconstchi}
{5.17}
{
     \frac{U}{1 - U \chi_0(0)} 
     = U + U^2 \chi_{\rm s}^{(0)}(0) > U . 
     }

Further, since $\Gamma_U^{(0)}$ becomes constant 
in the simplification mentioned above, 
it affects only $s$-wave pairing. 
Hence, we use the expression of $\Gamma_U^{(0)}$ 
for singlet pairing \eq.{eq:GammaU0singlet}. 
From eqs.~\refeq{eq:Gammac_plus_Gammaph_in_theta} 
and \refeq{eq:PhMInt_factor}, 
we obtain 
\Equation{eq:N0Gamma}
{5.18}
{
     \begin{split} 
     \Gamma_{\sigma \sigma'}(\vq,\omega) N(0) 
     & = 
         \frac{\mu_{\rm C}}{1 - \mu_{\rm C}^2} 
       + \frac{\mu_{\rm C}^2}{1 - \mu_{\rm C}} 
       + \frac{{\tilde F}(\vq)}{2 ( 1 + \mu_{\rm C}) } 
     \\
     & \hsp{4} 
       - \frac{g N(0) }{1 + \mu_{\rm C}} 
           {\tilde F}(\vq) \, 
           \theta ( \omega_{\rm c} - |\omega| ) , 
     \end{split}
     }
where we have defined a parameter $g$ by ${\bar \alpha} = 2 g N(0)$ and 
\Equation{eq:tildeFdef}
{5.19}
{
     {\tilde F}(\vq) 
     \equiv 
     \frac{{\tilde q}_{s}^2 }{ |\vq|^2 + {\tilde q}_{s}^2 } 
     }
with 
\Equation{eq:tildeqs_chi0const}
{5.20}
{
     {\tilde q}_{\rm s}^2 
     = \frac{q_{\rm s}^2}{1 + \mu_{\rm C}} 
     = \frac{q_{\rm TF}^2}{1 + \mu_{\rm C}} . 
     }
We expand ${\tilde F}(\vq)$ and $\Delta(\vk)$ on the Fermi surface 
with respect to the momentum components parallel to the Fermi surface 
as in eqs.~\refeq{eq:Veffexpand} and \refeq{eq:Deltaalphadef}. 
We define the expansion factors $C_{\alpha}$ of ${\tilde F}$ by 
\Equation{eq:tildeFexpand}
{5.21}
{
     {\tilde F}(\vk - \vk') 
     = 
     \sum_{\alpha} C_{\alpha} 
          \gamma_{\alpha}(k_{\parallel}) 
          \gamma_{\alpha}(k'_{\parallel}) . 
     }
For convenience, 
we write $\alpha = 0$ for $s$-wave pairing, 
while $\alpha \ne 0$ for anisotropic pairing, 
from now on.

Therefore, from \eq.{eq:N0Gamma} 
and the replacement of $\theta ( \omega_{\rm c} - |\omega| )$ 
with 
$ \theta (\omega_{\rm c} - |\xi_{\vk}|) 
  \theta (\omega_{\rm c} - |\xi_{\vk'}|) $ 
that is adopted in \S5.1 and \S5.3, 
the $\alpha$ component of the effective interactions is 
written in the form 
\Equation{eq:Valpha}
{5.22}
{
     V_{\alpha} N(0) 
     = - \lambda_{\alpha} 
         \theta (\omega_{\rm c} - |\xi_{\vk}|) 
         \theta (\omega_{\rm c} - |\xi_{\vk'}|) 
       + \mu_{\alpha} , 
     }
where $\lambda_{\alpha}$ and $\mu_{\alpha}$ denote 
the dimensionless coupling constants 
of the phonon-mediated pairing interaction 
and the repulsive Coulomb interaction, 
for ``$\alpha$-wave'' superconductivity, 
which are expressed as follows. 
The constant $\lambda_{\alpha}$ is expressed as 
\Equation{eq:lambdaalpha}
{5.23}
{
     \lambda_{\alpha} 
       = \frac{gN(0)}{1 + \mu_{\rm C}} C_{\alpha} , 
     }
for any value of $\alpha$. 
In contrast, the constant $\mu_\alpha$ has different forms 
depending on $\alpha = 0$ or $\alpha \ne 0$. 
For $s$-wave pairing ($\alpha = 0$), 
we obtain 
\Equation{eq:muCast0}
{5.24}
{
     \mu_{0}
       = 
            \frac{\mu_{\rm C}}{1 - \mu_{\rm C}^2} 
          + \frac{\mu_{\rm C}^2}{1 - \mu_{\rm C}} 
          + \frac{C_0}{2 ( 1 + \mu_{\rm C} )} , 
     }
while 
for anisotropic ``$\alpha$-wave'' pairing ($\alpha \ne 0$) 
\Equation{eq:muCastalpha}
{5.25}
{
     \mu_{\alpha} 
        = \frac{C_{\alpha}}{2 ( 1 + \mu_{\rm C} )} . 
     }
For the retardation effect, 
the effective Coulomb parameter $\mu_{\alpha}^{*}$ 
becomes smaller than $\mu_{\alpha}$ 
as 
\Equation{eq:muCast}
{5.26}
{
     \mu_{\alpha}^{*} 
       = \frac{ \mu_{\alpha} } 
              { 1 + \mu_{\alpha} 
                    \ln (W_{\rm C}/\omega_{\rm c}) } . 
     }
The derivation of \eq.{eq:muCast} is explained, for example, 
in Refs.~\citen{Sch83} and \citen{Shi03b}. 
The superconducting transition temperature 
for ``$\alpha$-wave'' pairing is estimated from the above parameters 
by the equation 
\Equation{eq:Tcalphawave}
{5.27}
{
     T_{{\rm c}\alpha} 
     = 1.13 \, \omega_{\rm c} \e^{- 1/{\tilde \lambda_{\alpha}}} , 
     }
where ${\tilde \lambda}_{\alpha} 
= \lambda_{\alpha} - \mu_{\alpha}^{*}$. 
As argued below \eq.{eq:Deltaalphadef}, 
the real transition temperature $\Tc$ is the highest one 
among all $T_{{\rm c}\alpha}$'s.

We obtain the isotope effect coefficient 
\Equation{eq:ie_alpha}
{5.28}
{
     \alpha_{\rm IE} = \frac{1}{2} 
          \left [ 1 - \Bigl ( 
               \frac{\mu_{\alpha}^{*}}
                    { \tilde \lambda_{\alpha} }
               \Bigr )^2 \right ] , 
     }
putting $\lambda_{\rm sf}^{\alpha} = 0$ in \eq.{eq:Tcalpha}. 
The reverse isotope effect $\alpha_{\rm IE} < 0$ occurs 
when $\mu_{\alpha}^{*} > {\tilde \lambda}_{\alpha}$, 
{\it i.e.}, $2 > \lambda_{\alpha}/\mu_{\alpha}^{*}$.

From eqs.~\refeq{eq:lambdaalpha} and~\refeq{eq:muCastalpha}, 
we have $\lambda_{\alpha}/\mu_{\alpha} = 2 g N(0)$ 
for anisotropic pairing. 
Hence, for $\alpha_{\rm IE} < 0$ to occur 
in anisotropic superconductors, 
the inequality $1 > gN(0)$ needs to hold, 
because 
$ 2 > \lambda_{\alpha}/\mu_{\alpha}^{*} 
    > \lambda_{\alpha}/\mu_{\alpha} > 2 g N(0)$. 
However, for such weak coupling, the transition temperature 
becomes extremely low so that it is not practically observable. 
If we take into account repulsive contribution to 
$\mu_{\alpha}$ from $\Gamma_U^{(0)}$, 
which modifies \eq.{eq:muCastalpha}, 
the reverse isotope effect $\alpha_{\rm IE} < 0$ is possible 
even in the present situation~\cite{Shi03b}.

In \eq.{eq:muCast}, the Coulomb cutoff energy $W_{\rm C}$ 
is usually of the order of the band width $W$. 
More precisely, it is defined by 
\Equation{eq:WCprecise}
{5.29}
{
     \ln W_{\rm C} 
       \equiv \frac{1}{2} 
       \Bigl [ 
           \ln \bigl ( W - \epsilon_{\rm F} \bigr ) 
         + \ln \epsilon_{\rm F} 
       \Bigr ] 
     }
as is easily verified from the derivation of \eq.{eq:muCast}, 
where $W$ and $\epsilon_{\rm F}$ denote 
the band width and the chemical potential (Fermi energy) 
measured from the bottom of the electron band. 
Equation~\refeq{eq:WCprecise} is written as 
$W_{\rm C} = \sqrt{ \epsilon_{\rm F} ( W - \epsilon_{\rm F} ) }$. 
Hence, it is easily proved that $W_{\rm C} \leq W/2$, 
where the equal sign holds when $\epsilon_{\rm F} = W/2$. 
In particular, when the charge carrier density is very small 
as in semi-metals, 
that is, $\epsilon_{\rm F} \ll W$, 
we obtain $W_{\rm C} \approx \sqrt{W \epsilon_{\rm F}} \ll W $.

%%% 5.5 
\subsection{Dimensionality and anisotropic pairing}

As we could see in eqs.~\refeq{eq:N0Gamma} - 
\refeq{eq:tildeFexpand}, 
the anisotropic pairing interactions can dominate 
when ${\tilde q}_{\rm s}$ is small 
and $C_{\alpha \ne 0}$ is near $C_0$. 
Since ${\tilde q}_{\rm s}$ is proportional to $q_{\rm TF}$ 
as seen in \eq.{eq:tildeqs_chi0const}, 
small $q_{\rm TF}$ favors anisotropic pairing. 
In systems with isotropic Fermi surfaces, 
we have 
\Equation{eq:Finalphas}
{5.30}
{
     {\tilde F}(\vq) 
     = \frac{ {\tilde q}_{\rm s}^2/2 \kF^2 } 
            { 1 - \cos \theta 
                +  {\tilde q}_{\rm s}^2/2 \kF^2 }
     = \frac{ \alpha_{\rm s} - 1 } 
            { \alpha_{\rm s} - \cos \theta } , 
     }
where $\alpha_{\rm s} = 1 + {\tilde q}_{\rm s}^2/2\kF^2 
= 1 + q_{\rm TF}^2/2 \kF^2 (1 + \mu_{\rm C}) $, 
and $\theta$ denotes the angle between $\vk$ and $\vk'$, 
since $\vq = \vk - \vk'$. 
Thus, it is found that 
the scaled parameters $q_{\rm TF}^2/2 \kF^2$ and $\mu_{\rm C}$ 
are the essential parameters 
in the sense that they determine the property 
of the function ${\tilde F}(\vq)$.

First, we discuss the three dimensional system. 
In systems with spherically symmetric Fermi surfaces~\cite{Shi02a}, 
we obtain 
\Equation{eq:qTFover2kFin3D}
{5.31}
{
     \frac{q_{\rm TF}^2}{2 \kF^2}
     = \frac{2}{\pi(3 \pi^2)^{1/3} n^{1/3} }
       \frac{m_{\rm b}}{m} \frac{a}{a_{\rm B}}
     \approx 
       \frac{1}{n^{1/3}} \frac{m_{\rm b}}{m} 
       \frac{a}{2.6 [\rm \AA]} 
     }
where $m$, $m_{\rm b}$, $a_{\rm B}$ and $n$ denote 
the bare electron mass, the band effective mass, 
the Borh radius $a_{\rm B} = \hbar^2/me^2 \approx 0.5293 {\rm \AA}$, 
and the electron (or charge carrier of $\pm e$) density per a site, 
respectively. 
Therefore, as the lattice constant $a$ increases, 
the ratio $q_{\rm TF}^2/2 \kF^2$ increases, 
which implies stronger screening.

We obtain a qualitatively different result 
in the quasi-two-dimensional system. 
In systems with cylindrically symmetric Fermi surfaces~\cite{Shi02b}, 
we obtain 
\Equation{eq:qTFover2kFin2D}
{5.32}
{
     \frac{q_{\rm TF}^2}{2 \kF^2}
     = \frac{1}{\pi n} 
       \frac{m_{\rm b}}{m} \frac{a^2}{a_{\rm B} d}
     \approx 
       \frac{1}{n} \frac{m_{\rm b}}{m} 
       \frac{a^2}{1.7 {[\rm \AA]} \times d} , 
     }
where $a$ and $d$ denote the lattice constant in the layer 
and the layer interval. 
As pointed out in our previous paper~\cite{Shi02b}, 
as the layer interval $d$ increases, 
the ratio $q_{\rm TF}^2/2 \kF^2$ decreases, 
which implies weaker screening, 
in contrast to the result in the three-dimensional systems 
mentioned above.

%%% 6. 
\section{Examples of the Application}

Purpose of this section is to show examples of the application of 
the effective Hamiltonian derived in the previous section. 
For simplicity, we consider spherical symmetric systems. 
We have already examined a spherical symmetric system 
in our previous paper~\cite{Shi02a}, 
but the present treatment is based on a more basic model. 
Consequently, the corrections due to $U$ are taken into account 
in more detail. 
The transition temperatures obtained in this section should be taken 
as crude estimations, because the model is simplified. 
Nevertheless, the qualitative results obtained here 
reflect essential aspects of the present mechanism.

From the symmetry, we take the basis function as 
$\gamma_{\alpha}({\hat \vk}) = Y_{lm}(\theta,\varphi)$ 
with $\alpha = (l,m)$. 
Therefore, 
\Equation{eq:tildeF_and_gap_expand_in_Ylm}
{6.1}
{
     \begin{split} 
     {\tilde F}(\vk - \vk') 
     & = 
     \sum_{l,m} C_{l} \, 
          [Y_{lm}(\theta,\varphi)]^{*} \, 
          Y_{lm}(\theta',\varphi')
     \\
     & = 
     \sum_{l} (2l + 1) \, C_{l} \, P_l(\cos \theta_{\vk\vk'}) 
     \\
     \Delta ({\hat \vk}) 
     & = 
     \sum_{lm} \Delta_{lm} 
          Y_{lm}(\theta,\varphi) , 
     \end{split}
     }
where $\theta_{\vk\vk'}$ denotes the angle between $\vk$ and $\vk'$. 
Since $P_0(w) = 1$ and $P_1(w) = w$, 
we obtain from \eq.{eq:Finalphas} 
\Equation{eq:C0C1}
{6.2}
{
     \begin{split} 
     C_{0} & = 
          \int \frac{\d \Omega}{4 \pi} 
               \frac{\alpha_{\rm s} - 1} 
                    { \alpha_{\rm s} - \cos \theta }
     = - \frac{ \alpha_{\rm s} - 1 }{2} 
          \ln \left | \frac{ \alpha_{\rm s} - 1 }{ \alpha_{\rm s} + 1 } \right | 
     \\
     C_{1} & = 
          \int \frac{\d \Omega}{4 \pi} 
               \frac{(\alpha_{\rm s} - 1) \, \cos \theta } 
                    { \alpha_{\rm s} - \cos \theta }
     \\ 
     & = - (\alpha_{\rm s} - 1) 
          \left [ 
            1 + \frac{\alpha_{\rm s}}{2} 
                \ln \left | \frac{ \alpha_{\rm s} - 1 }{ \alpha_{\rm s} + 1 } 
                    \right | 
          \right ] , 
     \end{split}
     }
where $\d \Omega = \sin \theta \, \d \theta \, \d \varphi$. 
Considering the signs of the three quantity, 
${\tilde \lambda}_0$, 
${\tilde \lambda}_1$, and 
${\tilde \lambda}_0 - {\tilde \lambda}_1$, 
we obtain the phase diagrams as shown in the following subsections.

%%%%%%%%%%%%%%%%%%%%%%%%%%%%%%%%%%%%%%%%%%%%%%%%%%%%%%%%%%%%%%%%%%%%%%%
%%  Fig.9                                                            %%
%%%%%%%%%%%%%%%%%%%%%%%%%%%%%%%%%%%%%%%%%%%%%%%%%%%%%%%%%%%%%%%%%%%%%%%
%% JPSJ %% 
\begin{figure}
%% JPSJ %% When you do not use epsf, activate the next line. 
%% \figureheight{7cm}
%% 
%% PR %% \begin{figure}[htb]
\begin{center}
%% FOR TWO COLUMN 
%% \leavevmode \epsfxsize=6cm  
%% 
%% FOR ONE COLUMN 
%% \leavevmode \epsfxsize=10cm  
%% JPSJ2 %% 
\vspace{2\baselineskip}
\begin{tabular}{c}
\includegraphics[width=6.5cm]{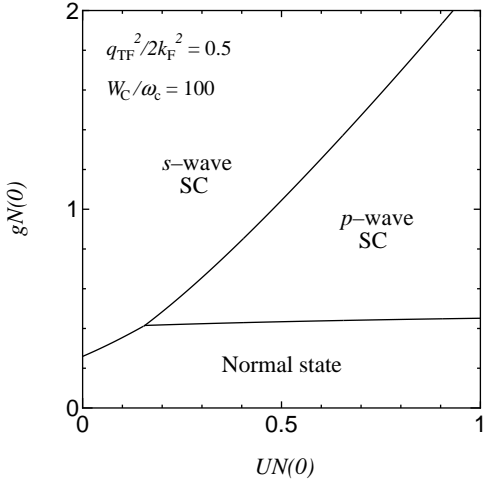}
\\[-20pt]
\end{tabular}
\vspace{\baselineskip}
%% PR %% \leavevmode \epsfxsize=7cm  
%% PR %% \epsfbox{fig09.eps}
\end{center}
\caption{
The phase diagram on the plane of the on-site Coulomb energy $U$ 
and the effective coupling constant $g$ 
of the phonon-mediated pairing interaction, 
in the spherical symmetric system. 
The parameters are chosen as 
$q_{\rm TF}^2/2 \kF^2 = 0.5$ and $W_{\rm C}/\omega_{\rm c} = 100$. 
The abbreviation SC stands for superconductivity. 
} 
\label{fig:3Dphasediagram01}
\end{figure}
%%%%%%%%%%%%%%%%%%%%%%%%%%%%%%%%%%%%%%%%%%%%%%%%%%%%%%%%%%%%%%%%%%%%%%%

%%%%%%%%%%%%%%%%%%%%%%%%%%%%%%%%%%%%%%%%%%%%%%%%%%%%%%%%%%%%%%%%%%%%%%%
%%  Fig.10                                                           %%
%%%%%%%%%%%%%%%%%%%%%%%%%%%%%%%%%%%%%%%%%%%%%%%%%%%%%%%%%%%%%%%%%%%%%%%
%% JPSJ %% 
\begin{figure}
%% JPSJ %% When you do not use epsf, activate the next line. 
%% \figureheight{7cm}
%% 
%% PR %% \begin{figure}[htb]
\begin{center}
%% FOR TWO COLUMN 
%% \leavevmode \epsfxsize=6cm  
%% 
%% FOR ONE COLUMN 
%% \leavevmode \epsfxsize=10cm  
%% JPSJ2 %% 
\vspace{2\baselineskip}
\begin{tabular}{c}
\includegraphics[width=6.5cm]{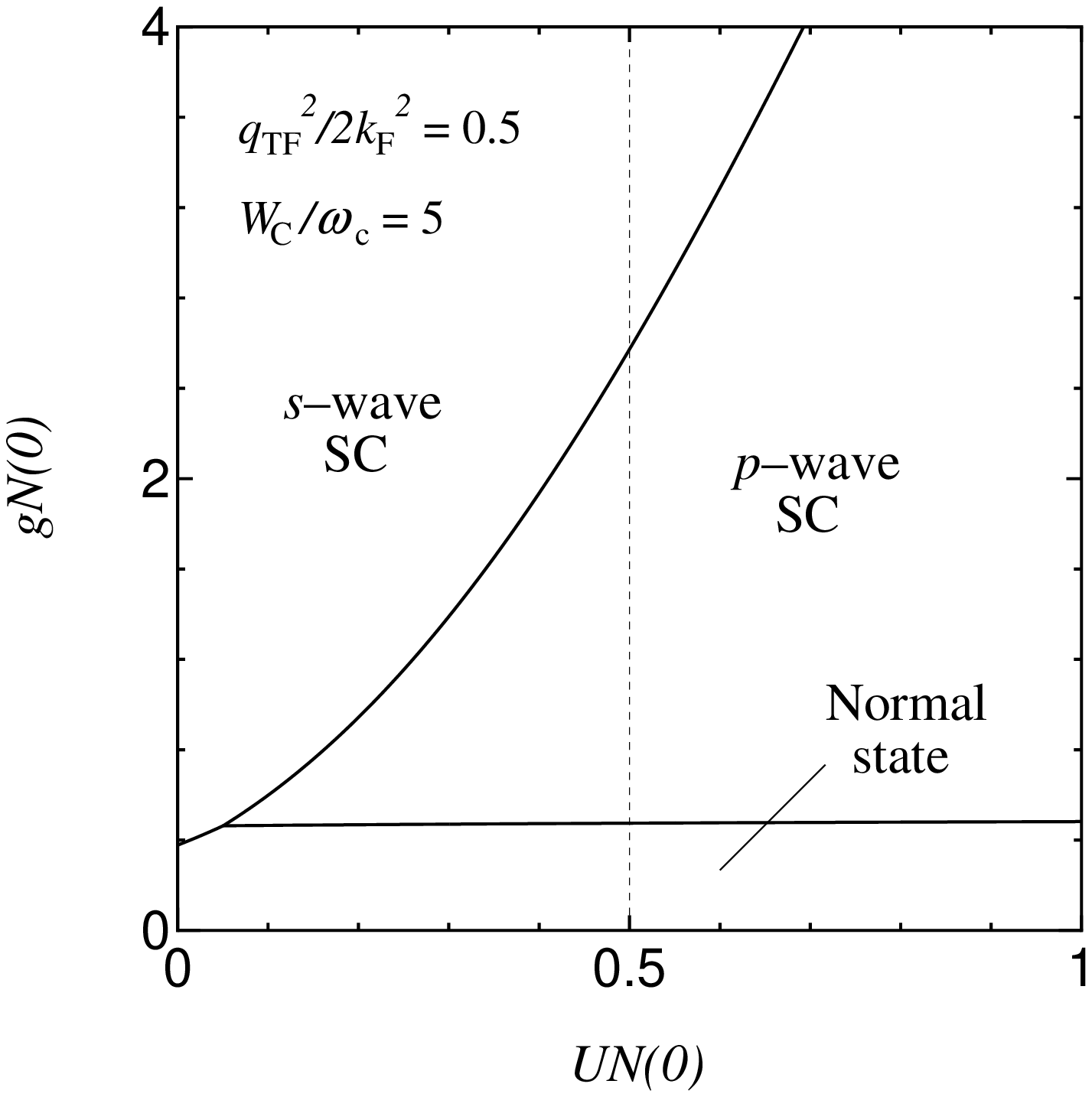}
\\[-20pt]
\end{tabular}
\vspace{\baselineskip}
%% PR %% \leavevmode \epsfxsize=7cm  
%% PR %% \epsfbox{fig10.eps}
\end{center}
\caption{
The phase diagram for a narrow band case, 
in which we set 
$q_{\rm TF}^2/2 \kF^2 = 0.5$ and $W_{\rm C}/\omega_{\rm c} = 5$. 
Along the vertical thin dotted line, 
the transition temperatures 
are plotted in Fig.~\ref{fig:3D_narrowband_Tc}. 
} 
\label{fig:3Dphasediagram_narrowband}
\end{figure}
%%%%%%%%%%%%%%%%%%%%%%%%%%%%%%%%%%%%%%%%%%%%%%%%%%%%%%%%%%%%%%%%%%%%%%%

%%%%%%%%%%%%%%%%%%%%%%%%%%%%%%%%%%%%%%%%%%%%%%%%%%%%%%%%%%%%%%%%%%%%%%%
%%  Fig.11                                                           %%
%%%%%%%%%%%%%%%%%%%%%%%%%%%%%%%%%%%%%%%%%%%%%%%%%%%%%%%%%%%%%%%%%%%%%%%
%% JPSJ %% 
\begin{figure}
%% JPSJ %% When you do not use epsf, activate the next line. 
%% \figureheight{7cm}
%% 
%% PR %% \begin{figure}[htb]
\begin{center}
%% FOR TWO COLUMN 
%% \leavevmode \epsfxsize=6cm  
%% 
%% FOR ONE COLUMN 
%% \leavevmode \epsfxsize=10cm  
%% JPSJ2 %% 
\vspace{2\baselineskip}
\begin{tabular}{c}
\includegraphics[width=6.5cm]{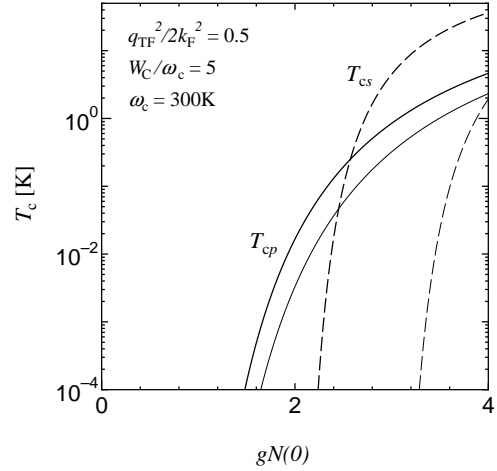}
\\[-20pt]
\end{tabular}
\vspace{\baselineskip}
%% PR %% \leavevmode \epsfxsize=7cm  
%% PR %% \epsfbox{fig11.eps}
\end{center}
\caption{
The transition temperatures as functions of the effective constant $g$ 
of the phonon-mediated interaction 
for the narrow band case, in which we set 
$q_{\rm TF}^2/2 \kF^2 = 0.5$, $W_{\rm C}/\omega_{\rm c} = 5$. 
We have put $\omega_{\rm c} = 300$~K as an example. 
The solid and dashed lines show the transition temperatures 
$T_{{\rm c}p}$ and $T_{{\rm c}s}$ 
of the $p$-wave and $s$-wave states for $UN(0) = 0.5$. 
For each value of $gN(0)$, the higher one of 
$T_{{\rm c}p}$ and $T_{{\rm c}s}$ is the final result of $\Tc$, 
{\it i.e.}, the physical $\Tc$. 
The thin solid and dashed lines show those for $UN(0) = 0.7$. 
} 
\label{fig:3D_narrowband_Tc}
\end{figure}
%%%%%%%%%%%%%%%%%%%%%%%%%%%%%%%%%%%%%%%%%%%%%%%%%%%%%%%%%%%%%%%%%%%%%%%

%%%%%%%%%%%%%%%%%%%%%%%%%%%%%%%%%%%%%%%%%%%%%%%%%%%%%%%%%%%%%%%%%%%%%%%
%%  Fig.12                                                           %%
%%%%%%%%%%%%%%%%%%%%%%%%%%%%%%%%%%%%%%%%%%%%%%%%%%%%%%%%%%%%%%%%%%%%%%%
%% JPSJ %% 
\begin{figure}
%% JPSJ %% When you do not use epsf, activate the next line. 
%% \figureheight{7cm}
%% 
%% PR %% \begin{figure}[htb]
\begin{center}
%% FOR TWO COLUMN 
%% \leavevmode \epsfxsize=6cm  
%% 
%% FOR ONE COLUMN 
%% \leavevmode \epsfxsize=10cm  
%% JPSJ2 %% 
\vspace{2\baselineskip}
\begin{tabular}{c}
\includegraphics[width=6.5cm]{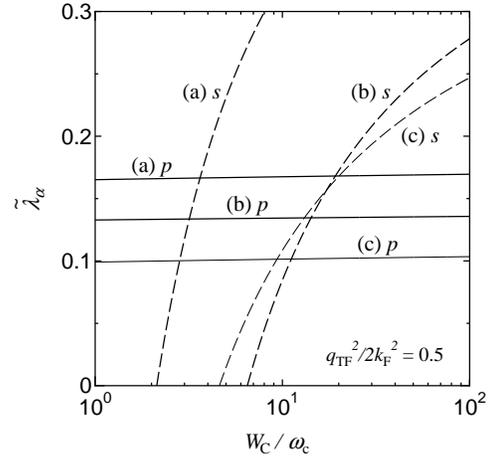}
\\[-20pt]
\end{tabular}
\vspace{\baselineskip}
%% PR %% \leavevmode \epsfxsize=7cm  
%% PR %% \epsfbox{fig12.eps}
\end{center}
\caption{
The dimensionless coupling constants ${\tilde \lambda}_{\alpha}$ 
as functions of $W_{\rm C}/\omega_{\rm c}$. 
The solid and dashed lines denote 
${\tilde \lambda}_{p}$ and ${\tilde \lambda}_{s}$, respectively. 
The parameter values are taken as 
(a) $gN(0) = 3$, $UN(0) = 0.5$, 
(b) $gN(0) = 3$, $UN(0) = 0.8$, 
and (c) $gN(0) = 2$, $UN(0) = 0.5$. 
} 
\label{fig:3D_WComegacdep}
\end{figure}
%%%%%%%%%%%%%%%%%%%%%%%%%%%%%%%%%%%%%%%%%%%%%%%%%%%%%%%%%%%%%%%%%%%%%%%

%%% 6.1 
\subsection{Typical parameter values for ordinary metals}

Figure~\ref{fig:3Dphasediagram01} shows the phase diagram for 
the set of parameter values $W_{\rm C}/\omega_{\rm c} = 100$ and 
$q_{\rm TF}^2/2 \kF^2 = 0.5$. 
The formar ratio $W_{\rm C}/\omega_{\rm c} = 100$ is 
typical of most metallic systems 
since $W_{\rm C} \sim W$ and $\omega_{\rm c} \sim \omegaD$. 
The latter ratio $q_{\rm TF}^2/2 \kF^2 = 0.5$ implies 
the Thomas-Fermi screening length 
$\lambda_{\rm TF} = 1/\kF \sim a/\pi$, 
which seems typical of metals, too. 
Because at $\mu_{\rm C} = U_{\rm c} N(0) \approx 1$, 
the magnetic instability $\chi_{\rm s} \rightarrow \infty $ 
should occur, the results are plotted only for $UN(0) < 1$. 
It is found that the triplet state occurs for large $U$, 
which confirms one of our previous results~\cite{Shi02a}. 
It is also found that for large $g$, 
the $s$-wave state is more favored than the $p$-wave state.

Therefore, for the present set of parameter values, 
the resultant dimensionless coupling constant 
${\tilde \lambda_p}$ for the $p$-wave state is quite small, 
where it overcomes the $s$-wave state. 
For example, when $\mu_{\rm C} = UN(0) = 0.5$, 
the highest transition temperature to the $p$-wave state occurs 
near the phase boundary at $gN(0) \approx 1.05$. 
However, at $\mu_{\rm C} = 0.5$ and $gN(0) = 1.0$, 
we obtain ${\tilde \lambda}_0 \approx 0.031$ and 
${\tilde \lambda}_1 \approx 0.037$, 
and such a small value of ${\tilde \lambda}_1$ gives 
$T_{{\rm c}p} \sim 10^{-10}$K, 
which is too low to be observed in practice. 
For $gN(0) \gsim 1.05$, we obtain larger $\Tc$, 
but there the $s$-wave state occurs. 
This result is consistent with the experimental fact that 
spin-triplet superconductivity has not been observed 
in the ordinary metals.

%%% 6.2 
\subsection{Narrow band systems} 

Next, we examine systems in which the ratio 
$W_{\rm C}/\omega_{\rm c}$ is not so large. 
From \eq.{eq:WCprecise}, 
since $W_{\rm C}/\omega_{\rm c} \sim 
          \sqrt{ \epsilon_{\rm F} ( W - \epsilon_{\rm F} )} 
          / \omega_{\rm D}$, 
such a situation is realized, for example, 
in semi-metals with extremely small $\epsilon_{\rm F}$ 
and in narrow band systems with extremely small $W$. 
In such systems, 
since $\omega_{\rm c} \ll W$ or $\omega_{\rm c} \ll \epsilon_{\rm F}$ 
does not hold, 
the approximation $N(\epsilon) \approx N(0)$ is not correct 
quantitatively, 
but it is not essential for our purpose. 
The phase diagram for $q_{\rm TF}^2/2 \kF^2 = 0.5$ 
and $W_{\rm C}/\omega_{\rm c} = 5$ is presented 
in Fig.~\ref{fig:3Dphasediagram_narrowband}. 
Since $W \gsim 2 W_{\rm C} = 10 \omega_{\rm c} \sim 10 \omegaD$, 
if we put $\omegaD = 300$~K, we obtain $W \gsim 0.3$~eV. 
As the ratio $W_{\rm C}/\omega_{\rm c}$ becomes smaller, 
since the retardation effect becomes weaker, 
$s$-wave pairing is suppressed more effectively 
by the on-site Coulomb repulsion. 
As a result, the area for the $p$-wave state is enlarged 
in the phase diagram, and $T_{{\rm c}p}$ becomes large enough 
to be observed in contrast to the result 
in the previous subsection {\S 6.1}.

In Fig.~\ref{fig:3D_narrowband_Tc}, 
the transition temperatures for the parameter values of 
Fig.~\ref{fig:3Dphasediagram_narrowband} are plotted, 
where we have put $\omegaD = 300$~K as an example. 
For $UN(0) = 0.5$, it is found that 
the transition temperature $T_{{\rm c}p}$ 
to the triplet state could reach 0.3~K 
at $gN(0) \approx 2.6$, 
while for $gN(0) \gsim 2.6$, $s$-wave superconductivity occurs 
since $T_{{\rm c}s} > T_{{\rm c}p}$. 
In contrast, for $UN(0) = 0.7$, 
since $T_{{\rm c}p} > T_{{\rm c}s}$ upto $gN(0) \approx 4$, 
because $s$-wave pairing is strongly suppressed by the on-site $U$. 
Thus, $T_{{\rm c}p}$ could reach a rather larger value 2~K, 
where the triplet state occurs.

Figure~\ref{fig:3D_WComegacdep} shows 
the dimensionless coupling constants ${\tilde \lambda}_{\alpha}$ 
as functions of $W_{\rm C}/\omega_{\rm c}$. 
It is found that the $p$-wave state is favored 
for narrower band widths $W$ ($\sim 2W_{\rm C}$), 
because the on-site $U$ suppresses $s$-wave pairing more effectively. 
The coupling constants ${\tilde \lambda}_{p}$ for $p$-wave pairing 
depend on $W_{\rm C}/\omega_{\rm c}$ slightly. 
If we put $\omega_{\rm c} \approx \omega_{\rm D} = 300~{\rm K}$, 
we could estimate $T_{{\rm c}p}$ as 
$T_{{\rm c}p} = 0.015~{\rm K}$, 0.16~K, and 0.79~K, 
for ${\tilde \lambda}_p = 0.1$, 0.13, and 0.165, respectively. 
Therefore, for $q_{\rm TF}^2/2 \kF^2 = 0.5$, which gives 
$\lambda_{\rm TF} \sim 1/\kF \gsim a/\pi \sim 0.3 \times a$, 
the transition temperature $T_{{\rm c}p}$ 
can be high enough to be observed in practice, 
where the $s$-wave state is suppressed 
by strong on-site Coulomb repulsion. 
For this mechanism, the smallness of 
the ratio $W_{\rm C}/\omega_{\rm c}$ is essential.

%%%%%%%%%%%%%%%%%%%%%%%%%%%%%%%%%%%%%%%%%%%%%%%%%%%%%%%%%%%%%%%%%%%%%%%
%%  Fig.13                                                           %%
%%%%%%%%%%%%%%%%%%%%%%%%%%%%%%%%%%%%%%%%%%%%%%%%%%%%%%%%%%%%%%%%%%%%%%%
%% JPSJ %% 
\begin{figure}
%% JPSJ %% When you do not use epsf, activate the next line. 
%% \figureheight{7cm}
%% 
%% PR %% \begin{figure}[htb]
\begin{center}
%% FOR TWO COLUMN 
%% \leavevmode \epsfxsize=6cm  
%% 
%% FOR ONE COLUMN 
%% \leavevmode \epsfxsize=10cm  
%% JPSJ2 %% 
\vspace{2\baselineskip}
\begin{tabular}{c}
\includegraphics[width=6.5cm]{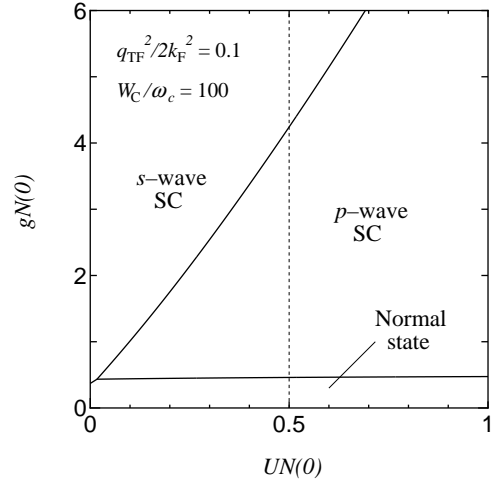}
\\[-20pt]
\end{tabular}
\vspace{\baselineskip}
%% PR %% \leavevmode \epsfxsize=7cm  
%% PR %% \epsfbox{fig13.eps}
\end{center}
\caption{
The phase diagram for a weak screening case, 
in which we set 
$q_{\rm TF}^2/2 \kF^2 = 0.1$ and $W_{\rm C}/\omega_{\rm c} = 100$. 
Along the vertical thin dotted line, the transition temperatures 
are plotted in Fig.~\ref{fig:3D_weakscreening_Tc}. 
}
\label{fig:3Dphasediagram_weakscreening}
\end{figure}
%%%%%%%%%%%%%%%%%%%%%%%%%%%%%%%%%%%%%%%%%%%%%%%%%%%%%%%%%%%%%%%%%%%%%%%

%%%%%%%%%%%%%%%%%%%%%%%%%%%%%%%%%%%%%%%%%%%%%%%%%%%%%%%%%%%%%%%%%%%%%%%
%%  Fig.14                                                           %%
%%%%%%%%%%%%%%%%%%%%%%%%%%%%%%%%%%%%%%%%%%%%%%%%%%%%%%%%%%%%%%%%%%%%%%%
%% JPSJ %% 
\begin{figure}
%% JPSJ %% When you do not use epsf, activate the next line. 
%% \figureheight{7cm}
%% 
%% PR %% \begin{figure}[htb]
\begin{center}
%% FOR TWO COLUMN 
%% \leavevmode \epsfxsize=6cm  
%% 
%% FOR ONE COLUMN 
%% \leavevmode \epsfxsize=10cm  
%% JPSJ2 %% 
\vspace{2\baselineskip}
\begin{tabular}{c}
\includegraphics[width=6.5cm]{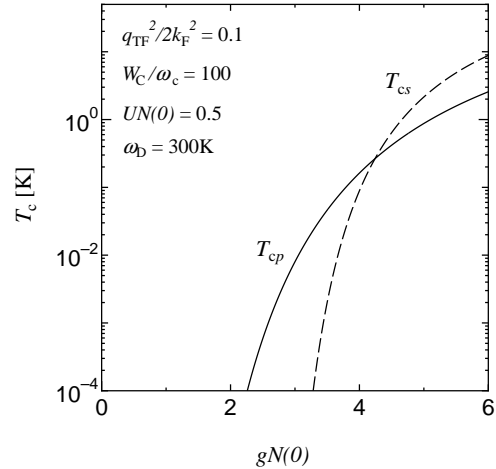}
\\[-20pt]
\end{tabular}
\vspace{\baselineskip}
%% PR %% \leavevmode \epsfxsize=7cm  
%% PR %% \epsfbox{fig14.eps}
\end{center}
\caption{
The transition temperatures as functions of the effective constant $g$ 
of the phonon-mediated interaction for a system with the weak screening, 
in which we set 
$q_{\rm TF}^2/2 \kF^2 = 0.1$, $W_{\rm C}/\omega_{\rm c} = 100$. 
We have put $\omega_{\rm c} = 300$~K as an example. 
The solid and dashed lines show the transition temperatures 
$T_{{\rm c}p}$ and $T_{{\rm c}s}$ 
for the $p$-wave and $s$-wave states. 
}
\label{fig:3D_weakscreening_Tc} 
\end{figure}
%%%%%%%%%%%%%%%%%%%%%%%%%%%%%%%%%%%%%%%%%%%%%%%%%%%%%%%%%%%%%%%%%%%%%%%

%%% 6.3 
\subsection{Systems with weak Coulomb screening} 

Next, we examine systems with weak Coulomb screening. 
As examined in our previous paper~\cite{Shi02b} 
and briefly discussed in \S5.5, 
the screening can be weak, 
for example in layered systems with sufficiently 
large intervals between the layers, 
although here we examine spherically symmetric systerms.

Now, let us examine the case with $q_{\rm TF}^2/2\kF^2 = 0.1$ and 
$W_{\rm C} / \omega_{\rm c} = 100$. 
For $q_{\rm TF}^2/2 \kF^2 = 0.1$, we obtain 
$\lambda_{\rm TF} = \sqrt{5}/\kF \gsim \sqrt{5}a/\pi \sim 0.7 \times a$. 
The short-range correlations lengthen the screening length 
by $\sqrt{1+\mu_{\rm C}}$ as shown in \eq.{eq:screeninglength}. 
In Fig.~\ref{fig:3Dphasediagram_weakscreening}, 
the area of the $p$-wave state is much larger than those in 
Figs.~\ref{fig:3Dphasediagram01}~and~\ref{fig:3Dphasediagram_narrowband}. 
Thus, we confirm that as the screening becomes weaker, 
the anisotropic pairing interactions increase, 
as proposed in our previous papers~\cite{Shi02a,Shi02b}.

In Fig.~\ref{fig:3D_weakscreening_Tc}, 
the transition termpartures are estimated 
for the parameter values of 
Fig.~\ref{fig:3Dphasediagram_weakscreening}. 
It is found that $T_{{\rm c}p}$ for the $p$-wave state 
could reach 0.3K, which are practically observable.

%%% 6.4 
\subsection{Narrow band systems with weak screening}

\begin{figure}
%% JPSJ %% When you do not use epsf, activate the next line. 
%% \figureheight{7cm}
%% 
%% PR %% \begin{figure}[htb]
\begin{center}
%% FOR TWO COLUMN 
%% \leavevmode \epsfxsize=6cm  
%% 
%% FOR ONE COLUMN 
%% \leavevmode \epsfxsize=10cm  
%% JPSJ2 %% 
\vspace{2\baselineskip}
\begin{tabular}{c}
\includegraphics[width=6.5cm]{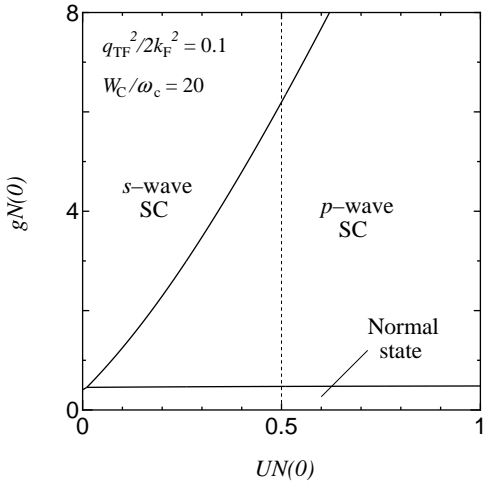}
\\[-20pt]
\end{tabular}
\vspace{\baselineskip}
%% PR %% \leavevmode \epsfxsize=7cm  
%% PR %% \epsfbox{fig15.eps}
\end{center}
\caption{
The phase diagram for a system with weak screening and 
a narrow electron band, 
in which we set 
$q_{\rm TF}^2/2 \kF^2 = 0.1$ and $W_{\rm C}/\omega_{\rm c} = 20$. 
Along the vertical thin dotted line, the transition temperatures 
are plotted in Fig.~\ref{fig:3D_narrowband_and_weakscreening_Tc}. 
}
\label{fig:3Dphasediagram_narrowband_and_weakscreening}
\end{figure}

\begin{figure}
%% JPSJ %% When you do not use epsf, activate the next line. 
%% \figureheight{7cm}
%% 
%% PR %% \begin{figure}[htb]
\begin{center}
%% FOR TWO COLUMN 
%% \leavevmode \epsfxsize=6cm  
%% 
%% FOR ONE COLUMN 
%% \leavevmode \epsfxsize=10cm  
%% JPSJ2 %% 
\vspace{2\baselineskip}
\begin{tabular}{c}
\includegraphics[width=6.5cm]{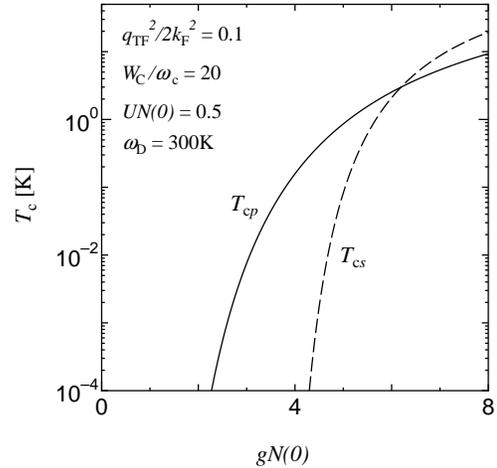}
\\[-20pt]
\end{tabular}
\vspace{\baselineskip}
%% PR %% \leavevmode \epsfxsize=7cm  
%% PR %% \epsfbox{fig16.eps}
\end{center}
\caption{
The transition temperatures as functions of 
the effective coupling constant $g$ 
of the phonon-mediated interaction 
for the system with the weak screening and the narrow electron band, 
in which we set 
$q_{\rm TF}^2/2 \kF^2 = 0.1$, $W_{\rm C}/\omega_{\rm c} = 20$. 
We have put $\omega_{\rm c} = 300$~K as an example. 
The solid and dashed lines show the transition temperatures 
$T_{{\rm c}p}$ and $T_{{\rm c}s}$ 
for the $p$-wave and $s$-wave states. 
}
\label{fig:3D_narrowband_and_weakscreening_Tc} 
\end{figure}

Lastly, we examine systems with weak Coulomb screening 
and narrow electron bands. 
We set parameters as 
$q_{\rm TF}^2/2 \kF^2 = 0.1$ and $W_{\rm C}/\omega_{\rm c} = 20$. 
The formar corresponds to the Thomas-Fermi screening length 
$\lambda_{\rm TF} \gsim 0.7 \times a$ 
as mentioned in the previous case. 
The latter corresponds to 
$W/\omega_{\rm D} \sim 2 W_{\rm C}/\omega_{\rm c} = 40$. 
If we put $\omega_{\rm D} = 200 \sim 400$~K as a typical value, 
we obtain $W = 0.8 \sim 1.6$~eV, which is realistic 
for the organic superconductors and the ruthenate superconductors. 
In Figs.~\ref{fig:3Dphasediagram_narrowband_and_weakscreening} 
and \ref{fig:3D_narrowband_and_weakscreening_Tc}, 
the phase diagram on the $U$-$g$ plane 
and the transition temperatures are shown. 
Comparing Figs.~\ref{fig:3Dphasediagram_weakscreening} 
and \ref{fig:3Dphasediagram_narrowband_and_weakscreening}, 
we confirm again that the area of the triplet state is enlarged 
when the electron band width becomes narrower. 
In Fig.~\ref{fig:3D_narrowband_and_weakscreening_Tc}, 
we find that the transition temperature $T_{{\rm c}p}$ 
of the triplet state 
can reach 3~K at $gN(0) \approx 6.2$, although for 
$gN(0) \gsim 6.2$, the $p$-wave state is hidden 
behind the $s$-wave state ($T_{{\rm c}s} > T_{{\rm c}p}$). 
We shall discuss the application of this result 
to the real compounds in the last section.

%%%%%%%%%%%%%%%%%%%%%%%%%%%%%%%%%%%%%%%%%%%%%%%%%%%%%%%%%%%%%%%%%%%%%%%
%%  Fig.15                                                           %%
%%%%%%%%%%%%%%%%%%%%%%%%%%%%%%%%%%%%%%%%%%%%%%%%%%%%%%%%%%%%%%%%%%%%%%%
%% JPSJ %% 
%%%%%%%%%%%%%%%%%%%%%%%%%%%%%%%%%%%%%%%%%%%%%%%%%%%%%%%%%%%%%%%%%%%%%%%

%%%%%%%%%%%%%%%%%%%%%%%%%%%%%%%%%%%%%%%%%%%%%%%%%%%%%%%%%%%%%%%%%%%%%%%
%%  Fig.16                                                           %%
%%%%%%%%%%%%%%%%%%%%%%%%%%%%%%%%%%%%%%%%%%%%%%%%%%%%%%%%%%%%%%%%%%%%%%%
%% JPSJ %% 
%%%%%%%%%%%%%%%%%%%%%%%%%%%%%%%%%%%%%%%%%%%%%%%%%%%%%%%%%%%%%%%%%%%%%%%

%%% 7. 
\section{Summary and Discussion}

%%% 7.1 
\subsection{Summary of the results} 

Now, we shall summarize the results. 
In \S 2, we have derived an effective Hamiltonian 
which is written in terms of 
the short-range (on-site) Coulomb interactions $U$ 
and the long-range Coulomb interactions $v_{\vq}$, 
from a basic Hamiltonian. 
On the basis of this effective Hamiltonian, 
we have examined the spin and charge fluctuations 
and screened Coulomb interactions (\S 3.1), 
considering the two-particle vertex part 
within an RPA extended to the model including both $U$ and $v_{\vq}$. 
The effective interactions between electrons are written 
in the simple forms as shown 
in eqs.~\refeq{eq:HCeffinsusceptibility} and 
\refeq{eq:HcfHsf}. 
We have derived an expression of the dielectric function 
and the screening length (\S 3.2). 
It was found that the screening length is lengthened 
by the on-site correlations due to $U$. 

In \S 3.3 and \S 3.4, 
we have examined the Coulomb screening corrections to 
the electron-phonon interactions and the phonon Green's function 
within the RPA consistent with those for 
the interactions between electrons. 
It was found that the sound velocity increases 
by the weakening of the screening 
due to the on-site correlations mentioned above.

In \S 4, we have examined superconductivity. 
We have derived a general form of the effective Hamiltonian 
which includes the interactions mediated by the phonons and 
those mediated by the spin and charge fluctuations, 
where the short and long-range parts of the Coulomb interactions 
and the screening effects are taken into account. 
It was found in \eq.{eq:Gammaefftheta} 
that the effective interaction is written 
in a summation of two terms, that is, $\Gamma_U^{(0)}$ and 
the renormalized phonon-mediated pairing interaction. 
Here, we call $\Gamma_U^{(0)}$ the Hubbard term, 
because it is nothing but the vertex part 
in the pure Hubbard model 
without the long-range Coulomb interaction $v_{\vq}$. 
In contrast, the latter term has the same form 
as the phonon-mediated interaction in the absence of 
the short-range Coulomb interaction $U$, 
but the prefactor and the phonon energy ${\tilde \omega}_{\vq}$ 
include the corrections due to~$U$.

In \S 5, we derived the effective interactions 
along the same way as the traditional weak coupling theory of 
superconductivity. 
In particular, for the system away from any magnetic instability, 
we derived the effective Hamiltonian, 
with which we have examined anisotropic superconductivity in \S6. 
The present model includes the models examined 
so far~\cite{Shi02a,Shi02b,Fri00,Cha00} 
in appropriate limits except the corrections due to $U$.

In \S 6, we have applied our theory to some ideal systems. 
The results are summarized as follows: 
(1)~The triplet superconductivity occurs in the system with 
strong on-site correlations and weak electron-phonon coupling; 
(2)~Spin-triplet pairing is favored when the band width is narrow 
and when Coulomb screening is weak~\cite{NoteWSCR}; 
(3)~The transition temperature of $p$-wave pairing is limited, 
because $s$-wave pairing is favored if the electron-phonon coupling 
is strong. 
In the most favorable system, 
the transition temperature $T_{{\rm c}p}$ 
can be only of the order of 1~K as a crude estimation. 
These results agree with our knowledge as follows: 
The triplet superconductivity has not been observed 
in the ordinary metals, 
while there are candidates in the heavy fermion superconductors, 
and the layered superconductors with large layer intervals, 
such as \TMTSFX {} and \SrRuO, 
but their $\Tc$'s are not so high.

%%% 7.2 
\subsection{Narrow band width and triplet pairing} 

As mentioned in the above summary, 
we have found in Figs.~\ref{fig:3Dphasediagram_narrowband} 
- \ref{fig:3D_WComegacdep} 
that $p$-wave pairing is favored 
in narrow band systems. 
This result can be explained as follows. 
The on-site Coulomb repulsion affects only $s$-wave pairing, 
but it has a limit itself. 
The effective Coulomb parameter $\mu_0$ satisfies an inequality 
\Equation{eq:limitofmuast}
{7.1}
{
     \mu_{0}^{*} 
       = \frac{\mu_0}{1 + \mu_0 \ln (W_{\rm C}/\omega_{\rm c})} 
       \lsim \frac{1}{1 + \ln (W_{\rm C}/\omega_{\rm c})} 
       \equiv \mu_{\rm max} , 
     }
where we have used $1 > U \chi_0(\vq,0) \gsim UN(0)$, 
which is the necessary condition 
for the absence of the magnetic long-range order 
within the RPA. 
The upper bound $\mu_{\rm max}$ 
is determined only by the ratio $W_{\rm C}/\omega_{\rm c}$, 
which is usually of the order of $W/\omega_{\rm D}$. 
Therefore, 
we obtain $\mu_{\rm max} \sim 0.18$, 0.30, and 0.38, 
for $W_{\rm C}/\omega_{\rm c} \sim 100$, 10, and 5, respectively. 
The phonon-mediated pairing interaction originally has 
the anisotropic components, such as $\lambda_{p}$, 
but they are much smaller than the isotropic component $\lambda_{s}$. 
For triplet pairing to occur only by phonon-mediated pairing interaction, 
it is needed that $s$-wave pairing is suppressed for some reason, 
and the dominancy of the coupling constants, 
$\lambda_s$ and $\lambda_p$, changes. 
From the results shown in Fig.~\ref{fig:3D_WComegacdep}, 
the value $\mu_{\rm max} \sim 0.18$ 
for $W_{\rm C}/\omega_{\rm c} \sim 100$ is 
too small to change the dominancy, 
when $\lambda_{\rm TF} \sim 1/\kF$. 
In contrast, 
the large value of $\mu_{\rm max} = 0.30 \sim 0.38$ 
for $W_{\rm C}/\omega_{\rm c} = 10 \sim 5$ 
can be large enough to suppress 
the $s$-wave state~\cite{Notelambda}.

%%% 7.3 
\subsection{Application to the heavy fermion systems}

The value of $\Tc$ obtained in Fig.~\ref{fig:3D_narrowband_Tc} 
is consistent with $\Tc \approx 0.5$~K in 
the heavy fermion ${\rm UPt_3}$ system~\cite{Sig91,Tou96}, 
if the set of the parameter values 
$q_{\rm TF}^2/2 \kF^2 = 0.5$ and $W_{\rm C}/\omega_{\rm c} = 5$ 
is regarded to be appropriate for this compound. 
In addition, as in ${\rm UGe_2}$~\cite{Sax00,Kir01}, 
if singlet pairing is suppressed by the internal exchange field 
due to the coexisting ferromagnetic long-range order, 
the $p$-wave transition temperature $T_{{\rm c}p}$ 
can be of the order of $1 \sim 2$~{\rm K}, 
which is consistent with the experimental value 
and our previous result~\cite{Shi02a,NoteHF}.

%%% 7.4 
\subsection{Conclusion and future study}

In conclusion, we have examined a model of the electron-phonon system 
with both the long-range and short-range Coulomb interactions, 
which is derived from a more basic model. 
In particular, we have examined superconductivity, 
and shown that the phonon-mediated interaction could 
induce spin-triplet pairing, 
when the Coulomb screening is relatively weak 
and the on-site Coulomb interaction is strong. 
The transition temperature can be high enough to be observed 
in practice, and similar to the experimental values 
in the candidates of the spin-triplet superconductors. 
In the future, more quantitative study on the isotope effect, 
in which the correlation effects and the strong coupling effects 
are taken into account, may give a direct evidence of 
the applicability of this mechanism.

For more precise estimations of $T_{{\rm c}p}$ in the application 
to \SrRuO {} compound, we need to take into account 
the two-dimensionality. 
We have examined a two-dimensional effective model 
in our previous paper~\cite{Shi02b}, 
and shown that a realistic value of $T_{{\rm c}p}$ 
can be reproduced. 
It is possible to derive the effective Hamiltonian 
from the present Hamiltonian 
with the interaction of \eq.{eq:Gammac_plus_Gammaph_in_theta}. 
The derivation will be presented in a separate paper~\cite{Shi04b}.

The application to the quasi-one-dimensional (Q1D) organic superconductor 
\TMTSFX {} is another interesting subject to study 
as discussed in \S1. 
In particular, Suginishi and the present author found 
by detailed calculations based on the present theory 
that specific features of the Q1D system play an important role 
in the pairing symmetry. 
It will also be presented in a separate paper~\cite{Sug04}.

\vspace{0.5\baselineskip}

\noindent
{\bf Acknowledgements}

This work was partly supported by 
a Grant-in-Aid for COE Research 
(No.13CE2002) of the Ministry of Education, 
Culture, Sports, Science and Technology of Japan.

%%%%%%%%%%%%%%%%%%%%%%%%%%%%%%%%%%%%%%%%%%%%%%%%%%%%%%%%%%%%%%%%%%%%%%%
%%  References                                                       %%
%%%%%%%%%%%%%%%%%%%%%%%%%%%%%%%%%%%%%%%%%%%%%%%%%%%%%%%%%%%%%%%%%%%%%%%

%%%%%%%%%%%%%%%%%%%%%%%%%%%%%%%%%%%%%%%%%%%%%%%%%%%%%%%%%%%%%%%%%%%%%%%


\begin{thebibliography}{99}
\bibitem{NoteWSCR}
  Here, we do not consider the system with extremely weak screening 
  so that such a system becomes an insulating Wigner lattice.
\bibitem{NoteORG}
  The results of the thermal conductivity in Ref.~\citen{Bel97} 
  suggest a full gap state, 
  while the NMR results in Ref.~\citen{Tak87} 
  suggest existence of the line nodes. 
  This discrepancy has been discussed 
  in Refs.~\citen{Shi00c} and~\citen{Che03}. 
  It has sometimes been claimed that sufficiently low temperatures 
  to prove the existence of line nodes were not reached 
  in the experiments in Ref.~\citen{Tak87}. 
  However, at present, there is not any explanation of 
  $T_1^{-1} \sim T^3$ for $\Tc \gsim T \gsim \Tc/2$ 
  without line nodes or a nearly line-node-like behavior~\cite{Shi00c} 
  of the gap function, as far as the author's knowledge.
\bibitem{NoteCEX}
  The interactions mediated by antiferromagnetic fluctuations include 
  attractive components which contribute to 
  triplet pairing~\cite{Sca86,Shi88,Shi89,Shi00e}.
\bibitem{NoteHe3}
  The ratios $T_{{\rm c}p}/\epsilon_F$ happen to be of the same order 
  in the ${\rm Sr_2RuO_4}$ system and the liquid ${}^3{\rm He}$. 
  One might consider that this suggests an analogy between these systems, 
  but actualy it is only accidental. 
  It is needless to say, the repulsions between electrons and that 
  between ${}^3{\rm He}$ atoms are different in their magnitudes and 
  momentum dependences.
\bibitem{NoteRTI}
  Note that \eq.{eq:chargespin} is rotational invariant.
\bibitem{NoteSEEonTc}
  The significance of the self-energy effects on $\Tc$ 
  in the spin fluctuation mechanism has been clarified 
  in Ref.~\citen{Shi89}.
\bibitem{Notelambda}
  For example, if we put $\Tc = 10$~K 
  and $\omega_{\rm D} = 300$~K, 
  we obtain $\lambda \approx 0.28$ 
  by the weak coupling formula of $\Tc$, 
  which is smaller than $\mu_{\rm max} = 0.30 \sim 0.38$.
\bibitem{NoteHF}
  For the heavy fermion compound ${\rm UPt_3}$, 
  the value $q_{\rm TF}^2/2 \kF^2 = 0.5$ may be considered to be too small, 
  because $q_{\rm TF}^2/2 \kF^2 \propto m_{\rm b}/m$ 
  in \eq.{eq:qTFover2kFin3D}. 
  If we use the effective mass $m^{*}$ of the heavy fermions 
  as $m_{\rm b}$, the factor $m_{\rm b}/m \sim 180$ 
  for ${\rm UPt_3}$~\cite{Vis87} 
  implies an extremely large value of $q_{\rm TF}^2/2 \kF^2$ 
  of the order of 100. 
  However, as discussed in \S1, 
  it is obviously an artifact of the Thomas-Fermi approximation 
  and the jellium model. 
  For a practical purpose, 
  it is more appropriate to take a smaller value 
  such as $q_{\rm TF}^2/2\kF^2 = 0.5$, 
  so that an appropriate value of the nearest neighbor Coulomb interaction 
  is reproduced.
\end{thebibliography}
\end{document}